\documentclass[12pt,preprint]{aastex}

\shortauthors{Bower et al.}
\shorttitle{Monitoring of the G2 Cloud}
\begin{document}

\newcommand\degd{\ifmmode^{\circ}\!\!\!.\,\else$^{\circ}\!\!\!.\,$\fi}
\newcommand{\etal}{{\it et al.\ }}
\newcommand{\uv}{(u,v)}
\newcommand{\rdm}{{\rm\ rad\ m^{-2}}}
\newcommand{\msuny}{{\rm\ M_{\sun}\ y^{-1}}}
\newcommand{\mylesssim}{\stackrel{\scriptstyle <}{\scriptstyle \sim}}
\newcommand{\lsim}{\stackrel{\scriptstyle <}{\scriptstyle \sim}}
\newcommand{\gsim}{\stackrel{\scriptstyle >}{\scriptstyle \sim}}
\newcommand{\sci}{Science}
\newcommand{\sgr}{PSR J1745-2900}
\newcommand{\sgra}{Sgr~A*}
\newcommand{\kms}{\ensuremath{{\rm km\,s}^{-1}}}
\newcommand{\masy}{\ensuremath{{\rm mas\,yr}^{-1}}}

\def\kbar{{\mathchar'26\mkern-9mu k}}
\def\totd{{\mathrm{d}}}


\title{Radio and Millimeter Monitoring of Sgr A*:  Spectrum, Variability, and Constraints on the G2 Encounter}

\author{
Geoffrey C.\ Bower,\altaffilmark{1}
Sera Markoff,\altaffilmark{2}
Jason Dexter,\altaffilmark{3}
Mark A. Gurwell,\altaffilmark{4}
James M. Moran,\altaffilmark{4}
Andreas Brunthaler,\altaffilmark{5}
Heino Falcke,\altaffilmark{5,6,7}
P. Chris Fragile,\altaffilmark{8}
Dipankar Maitra,\altaffilmark{9}
Dan Marrone,\altaffilmark{10}
Alison Peck,\altaffilmark{11}
Anthony Rushton,\altaffilmark{12,13}
Melvyn C.H. Wright\altaffilmark{14}
}

\altaffiltext{1}{Academia Sinica Institute of Astronomy and Astrophysics, 645 N. A'ohoku Place, Hilo, HI 96720, USA; gbower@asiaa.sinica.edu.tw}
\altaffiltext{2}{Anton Pannekoek Institute for Astronomy, University of Amsterdam, Science Park 904, 1098XH Amsterdam, The Netherlands}
\altaffiltext{3}{Max Planck Institute for Extraterrestrial Physics, Giessenbachstr. 1, 85748 Garching, Germany}
\altaffiltext{4}{Harvard-Smithsonian Center for Astrophysics, 60 Garden Street, Cambridge, MA 02138, USA}
\altaffiltext{5}{Max-Planck-Institut f\"ur Radioastronomie, Auf dem H\"ugel 69, D-53121 Bonn, Germany}
\altaffiltext{6}{Department of Astrophysics, Institute for Mathematics, Astrophysics and Particle Physics (IMAPP), Radboud University, PO Box 9010, 6500 GL Nijmegen, The Netherlands}
\altaffiltext{7}{ASTRON, P.O. Box 2, 7990 AA Dwingeloo, The Netherlands}
\altaffiltext{8}{Department of Physics and Astronomy, College of Charleston, Charleston, SC 29424, USA}
\altaffiltext{9}{Department of Physics and Astronomy, Wheaton College, Norton, MA 02766, USA}
\altaffiltext{10}{Steward Observatory, University of Arizona, 933 North Cherry Avenue, Tucson, AZ 85721, USA}
\altaffiltext{11}{National Radio Astronomy Observatory, 520 Edgemont Road, Charlottesville, VA 22903-2475, USA}
\altaffiltext{12}{Department of Physics, Astrophysics, University of Oxford, Keble Road, Oxford OX1 3RH, UK}
\altaffiltext{13}{School of Physics and Astronomy, University of Southampton, Highfield, Southampton SO17 1BJ, UK}
\altaffiltext{14}{Radio Astronomy Laboratory, University of California, Berkeley, CA 94720-3411, USA}

\begin{abstract}
  We report new observations with the Very Large Array, Atacama Large
  Millimeter Array, and Submillimeter Array at frequencies from 1.0 to
  355 GHz of the Galactic Center black hole, Sagittarius A*.  These
  observations were conducted between October 2012 and November 2014.
  While we see variability over the whole spectrum with an amplitude as
  large as a factor of 2 at millimeter wavelengths,
  we find no evidence for a change in the mean flux density or
  spectrum of Sgr A* that can be attributed to interaction with the G2
  source.  The absence of a bow shock at low frequencies is consistent
  with a cross-sectional area for G2 that is less than
  $2 \times 10^{29}$ cm$^2$.  This result fits with several model predictions
  including a magnetically arrested cloud, a pressure-confined stellar
  wind, and a stellar photosphere of a binary merger.  There is no
  evidence for enhanced accretion onto the black hole driving greater
  jet and/or accretion flow emission.  Finally, we measure the
  millimeter wavelength spectral index of Sgr A* to be flat; combined with previous
  measurements, this suggests that there is no spectral break between
  230 and 690 GHz.  
The emission region is thus likely in a transition
  between optically thick and thin 
  at these frequencies  and requires a mix of lepton
  distributions with varying temperatures consistent with
  stratification.
\end{abstract}

\keywords{black hole physics, accretion, galaxies:  jets, galaxies:  active, Galaxy:  center}

\section{Introduction}

Accretion provides the energy that powers radiation from the Galactic
Center black hole, Sagittarius A*
\citep{2010RvMP...82.3121G,2013CQGra..30x4003F}.  The VLT NIR
discovery of an infalling object onto \sgra, named G2, offers an
especially exciting possibility for probing the nature of that
accretion flow via potential interactions as the object goes through
periastron \citep{2012Natur.481...51G}.  
If it is a pressure-confined gas cloud,
the G2 object is estimated to
have a gas mass of $3 M_\earth$, which is comparable to the total
accretion flow mass within its closest approach.  Best estimates
indicate a highly eccentric orbit with a periastron to \sgra\
of $\sim 2000$ Schwarzschild radii ($ R_{\rm Sch}$), a mere $2\%$ of
the Bondi accretion radius, in early 2014
\citep{2013ApJ...774...44G,2013ApJ...773L..13P,2015ApJ...798..111P}.
The origin, structure, and fate of the object remain uncertain.  While
the VLT observations of the Br-$\gamma$ transition indicate that G2 is being disrupted, through
tidal processes and possibly through gas dynamical processes, Keck NIR
observations have shown a persistently compact $L^\prime$ continuum source
\citep{2014ApJ...796L...8W}.

These observational results have led to a
range of models, from diffuse clouds to dense, dusty stellar winds
surrounding stars
\citep{2012ApJ...755..155S,2012NatCo...3E1049M,2013ApJ...768..108S,2014ApJ...786L..12G}.
Both the nature of the accretion process and the timescale on which it
takes place depend sensitively on the density profile of the accretion
region and the initial density structure of the object, among other
factors \citep{2012ApJ...755..155S,2012ApJ...759..132A}. Proposed
models anticipate a steep increase in the accretion rate, potentially
by several orders of magnitude, and heightened variability on
timescales of months to decades \citep{2012ApJ...755..155S}.  The G2
encounter could shed light on the historic and episodic Galactic
Center flares seen through X-ray fluorescence
\citep{2004A&A...425L..49R}.  The processing of G2's material through
the accretion flow around \sgra\ presents a unique opportunity to
study the radial density and temperature structure of the flow,
filling in crucial gaps in our understanding of ultra-low-Eddington-luminosity
accretion.

The long-term monitoring at radio 
and millimeter wavelengths for \sgra\ demonstrates short-term variability
and long-term stability,
consistent with damped random walk evolution
\citep{2006ApJ...641..302M,2014MNRAS.442.2797D}.  Since the detection
of Sgr A* at shorter wavelengths, X-ray and NIR light curves have also
not shown any secular evolution
\citep{2009ApJ...691.1021D,2011ApJ...728...37D,2012ApJS..203...18W,2013ApJ...774...42N}.
Intensive campaigns to monitor \sgra\ over a wide range of wavelengths have
been launched since the discovery of G2 was announced, none of which
have detected significant changes in the flux density or activity of
\sgra\
\citep{2014IAUS..303..288A,2014ATel.6242....1H,2015ApJ...798L...6T,2014ApJ...793..120H},
although those searches did lead to the discovery of the Galactic
Center pulsar \sgr\ \citep{2013ATel.5006....1D,2013ApJ...770L..24K,2013ApJ...770L..23M,2013ApJ...775L..34R,2013Natur.501..391E,2013MNRAS.435L..29S}.

Among the predictions for G2 are strong increases in radio through
millimeter spectrum driven by two processes.
Increasing the mass
accretion rate at separations of a few Schwarzschild radii would be expected to lead
to enhanced emission from the accretion flow and/or jet  
\citep{1993A&A...278L...1F,2007MNRAS.379.1519M,2012ApJ...752L...1M}.  
The
timescale for this accretion event may be as short as the free-fall
time ($\sim 0.1$ yr) or as long as the viscous time
(0.1 -- 100 yr).  
Additionally, a strong precursor
effect has been predicted for low radio frequencies due to the
potential shock forming between the incoming object and the accretion
flow \citep{2012ApJ...757L..20N}.  In this paper, we present Karl
G. Jansky Very Large Array (VLA), Atacama Large Millimeter Array
(ALMA), and Submillimeter Array (SMA) observations that span the time
since the discovery of G2 through periastron.  In Section~\ref{sec:obs},
we present the observations and analysis techniques.  In
Section~\ref{sec:results}, we give the results, which show a marked
lack of any change in level of activity relative to historical variability.  In
Section~\ref{sec:discussion}, we discuss and summarize our
conclusions.

\section{Observations and Data Reduction \label{sec:obs}}

\subsection{VLA Observations}

VLA observations were carried out as part of a service
observing program on the advice of an ad hoc committee \citep[project code TOBS0006;][]{2014IAUS..303..327S} between late 2012 and mid-2014 
(Table~\ref{tab:vlaobs}) at frequencies between 1 and 41 GHz
(Table~\ref{tab:frequencytable}).  
Observations were obtained in 8 separate observing bands,
each with a total bandwidth of 2 GHz
with the exception of L band observations, which
had a total bandwidth of 1 GHz.  
For each band,
the correlator was configured with 16 spectral windows, each with 64 channels,
for a total of 1024 channels spanning 
the bandwidth of 2 GHz (1  GHz in the case of L band).
Observations were snapshots of
a few minutes' duration in each of the eight frequency bands.  Observations
of 3C 286 were obtained for absolute gain calibration.

Standard pipeline calibration techniques in CASA were carried out by
NRAO staff \citep{2014IAUS..303..327S}.  The pipeline produced calibrated
visibility files that we then analyzed with our own software.
Table~\ref{tab:vlaobs} summarizes beam sizes for Sgr A* for L and
Q band (1.5 and 40 GHz, respectively) from images obtained from all of the
data in each epoch.  Unlike the reduction described below, the images included
short baselines as a means of demonstrating the large beams that occurred.  
The ratio of Q to L band beam sizes does not exactly follow the expected
frequency scaling ratio, which is indicative of the poorly shaped beams from
these snapshot observations and frequency-dependent weighting in
the imaging stage.
We averaged the data in time into 30s intervals and in
frequency for each spectral window.  We then exported the data for
further analysis with our code.

We performed two kinds of analyses.  For the first, we extracted
flux densities for \sgra\ by averaging visibility amplitudes on baselines
longer than 50 $k\lambda$.  Past experience has shown that these long
baselines are sufficient to exclude the effects of extended structure
on the flux density of \sgra\ \citep[e.g.,][]{2004AJ....127.3399H}.  At high frequencies and in
extended configurations, we determine flux densities for 
\sgra\ that are consistent with those provided by \citet{2014IAUS..303..327S}.

For the second analysis, we used a visibility subtraction method to
determine deviations from the average flux density.  This method is
important for low frequency observations in compact configurations,
where the extended flux dominates that of \sgra.  These epochs
were reported to have upper limits as high as 15 Jy \citep[e.g.,][]{2013ATel.5153....1C}.
For each epoch, we gridded the visibilities for each spectral window
and then found individual grid cells for which there was an overlap
between the individual epoch and the remaining ensemble.  We
differenced each epoch's visibility grid against the grid derived from
the ensemble of visibilities (minus the particular epoch).  The
differential flux density was determined as a weighted sum of the
residual visibilities.  We applied a weighting scheme that scales as
the inverse of the $(u,v)$ distance.  That is, longer baselines have
greater weight than shorter baselines in order to bias against
contamination from extended structure.  Different weighting schemes do
not qualitatively alter our conclusions.  The scatter in the
differences determines the error.  We then averaged over each observing
band and, again, calculated the error based on the scatter in the
measurements.

This visibility subtraction scheme works well given the large number of data
sets and the fine frequency resolution of the observations.  The 
frequency resolution permits us to make comparisons that are not limited
by the steep spectral indices of some of the extended structures present
in the Galactic Center.  The large number of data sets gives good
overlap between extended and compact structures.  
The method is sensitive to the effect of pairs of observations with significant
overlap in $(u,v)$ coverage, i.e., those conducted at
the same sidereal time and in the same configuration.  In this case,
the method effectively differences those two epochs and does not provide
a difference with respect to the average.  This effect can be seen
in the epochs 20131129 and 20131229, in which the flux difference 
for one mirrors the other.  The apparent minimum in rms variability 
of the flux density excess at 5 GHz is the result of increasing variability
toward  high frequencies and decreasing accuracy of the method toward
low frequencies because of the smaller number of long baselines.

\begin{figure}[p!]
\includegraphics[]{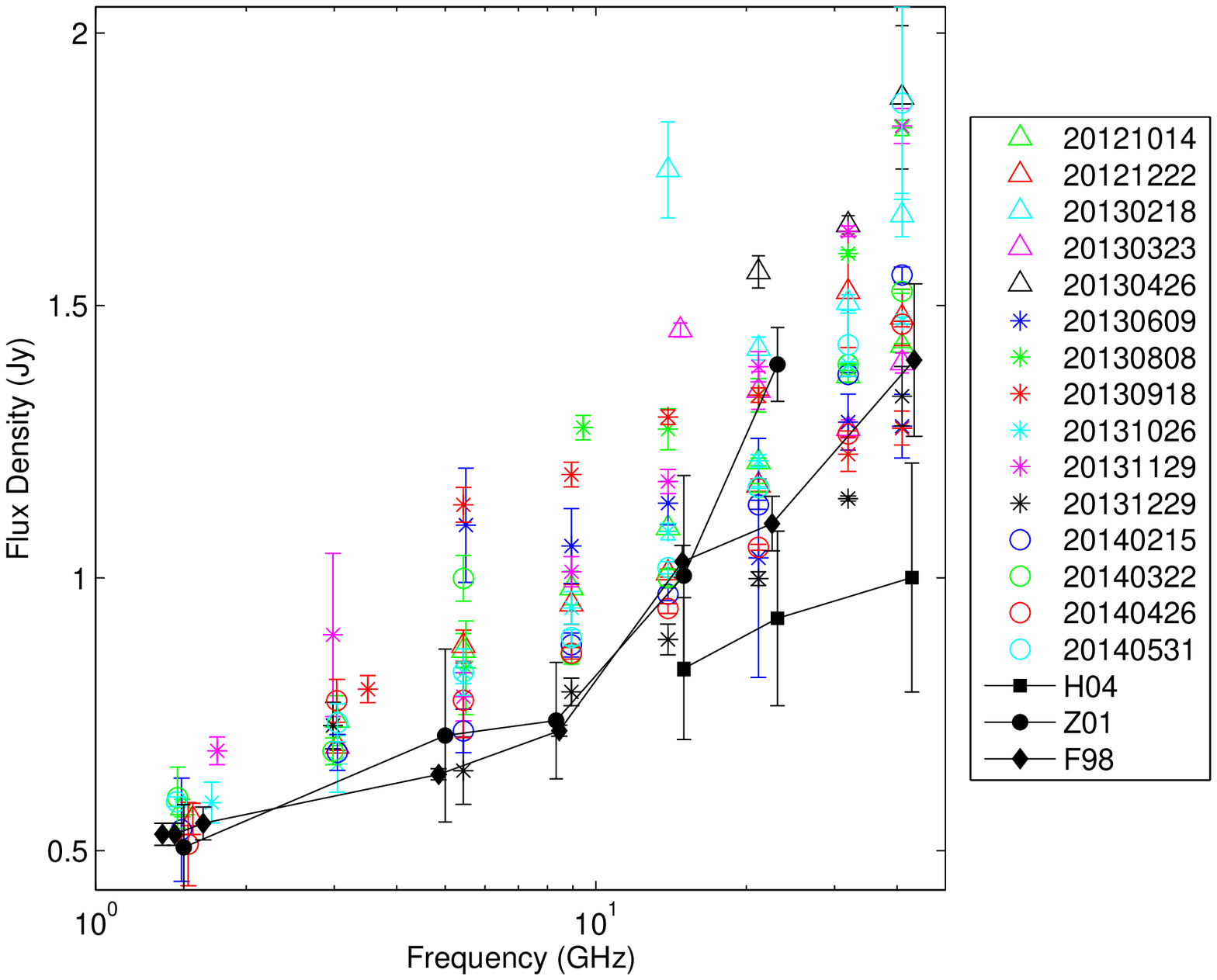}
\caption{
Spectrum of \sgra\ from VLA observations.  Absolute flux densities are determined using
long baseline data for epochs where it is available.
We also plot results from historical measurements of the flux density:  \citet[][F98]{1998ApJ...499..731F},
\citet[][Z01]{2001ApJ...547L..29Z}, and \citet[][H04]{2004AJ....127.3399H}.  The F98 
data consist of one measurement at a single epoch while Z01 and H04 represent averages over many years.
\label{fig:specvla}
}
\end{figure}

\begin{figure}[p!]
\includegraphics[]{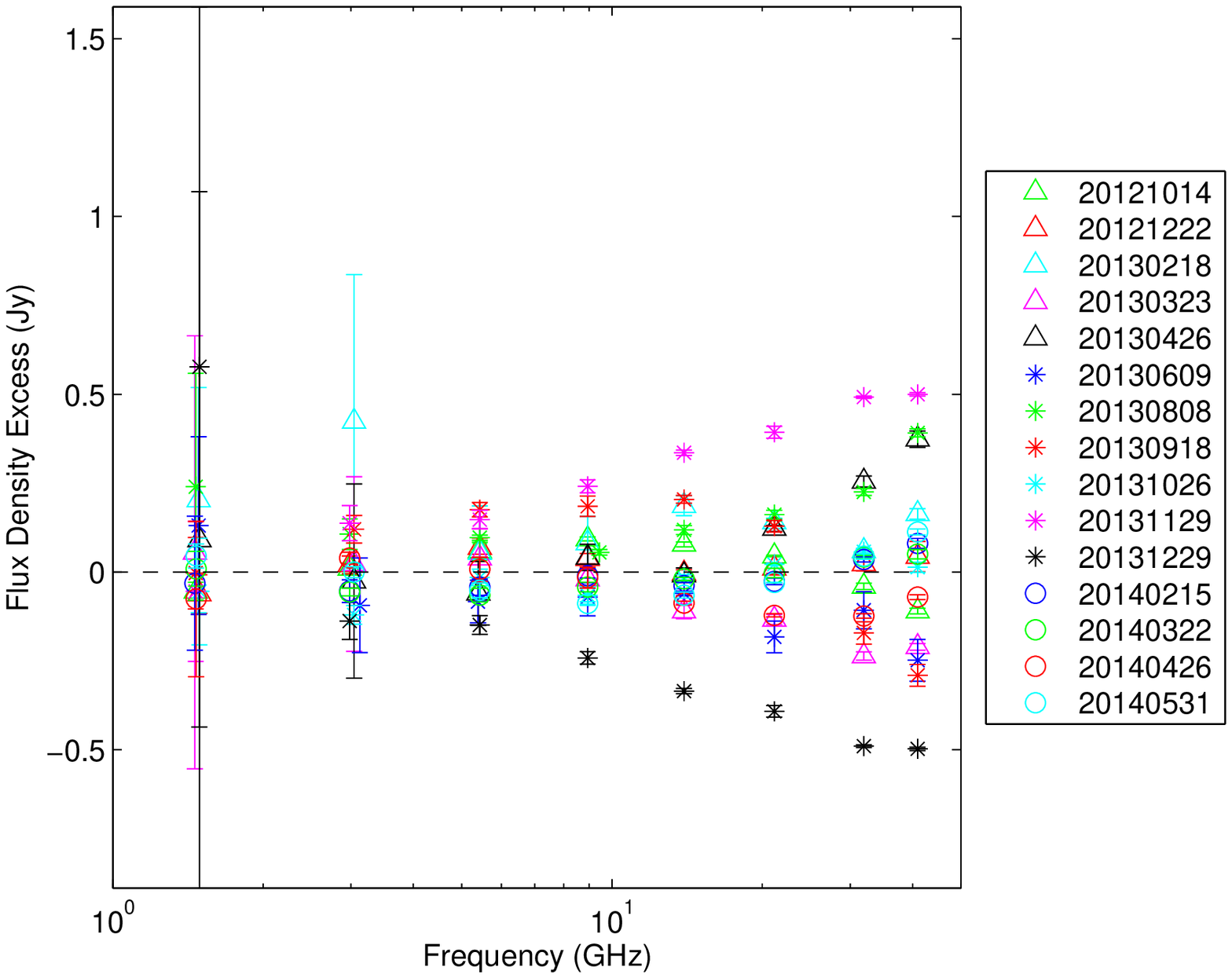}
\caption{
Differential spectrum of \sgra\ from all VLA epochs.  
\label{fig:diffspecvla}
}
\end{figure}

We summarize the mean flux densities and differential flux densities in
Table~\ref{tab:vlaresults}.
We show total intensity and differential spectra for \sgra\ in Figures~\ref{fig:specvla} and ~\ref{fig:diffspecvla}.  We plot the
total intensity spectrum alongside historical average spectra \citep{1998ApJ...499..731F,2001ApJ...547L..29Z,2004AJ....127.3399H}.

\subsection{ALMA Observations}

We carried out ALMA observations between mid-2013 and mid-2014
(Table~\ref{tab:almaobs}), over eight epochs, each with band 6 (230
GHz) and band 7 (345 GHz) receivers.  Details of the frequency setup
are presented in Table~\ref{tab:frequencytable}.  One spectral window
in each band was configured with higher spectral resolution in order
to permit detection of hydrogen recombination lines.  We will discuss
those results in a separate paper.  Observations were snapshots with
durations of a few minutes.  
The total flux density, corresponding to Stokes I, was produced by
summing the response of parallel handed cross-correlations of 
the  orthogonal linearly polarized feeds.

\begin{figure}[p!]
\includegraphics{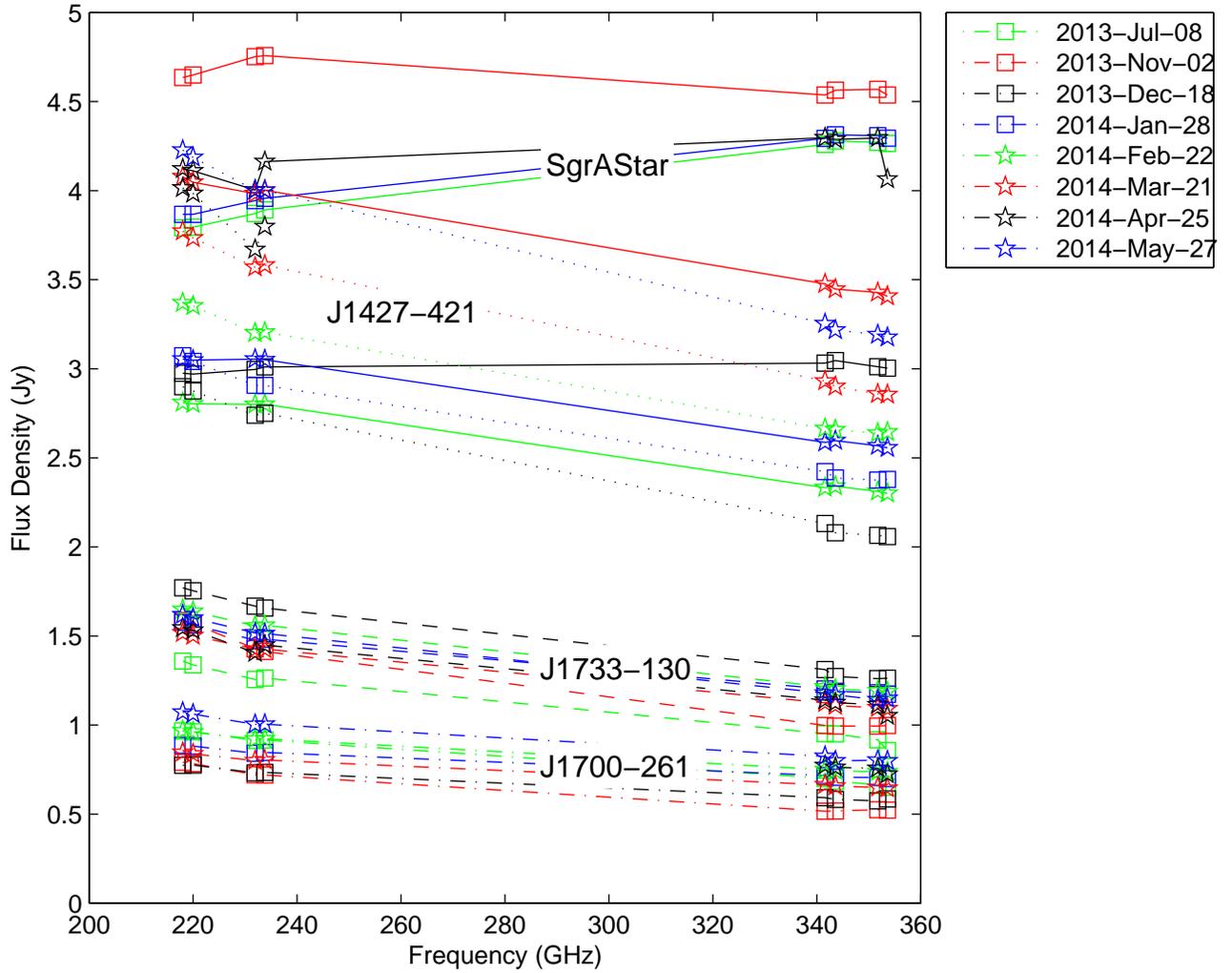}
\caption{
Spectrum of \sgra\ and calibrators from ALMA observations.
Lines connect spectra
for \sgra\ (solid), J1427-421 (dotted), J1733-130 (dashed), and J1700-261 (dash-dotted).  Symbols
and colors denote the observational epoch.
\label{fig:specalma}
}
\end{figure}

We set absolute amplitude calibration with observations of Titan and
Neptune and set bandpass gains with observations of the compact source
J1427-421.  All calibrator sources and \sgra\ were
phase-self-calibrated on timescales of a single integration (10 sec).
For \sgra, phase self-calibration solutions were obtained only for
baselines longer than $50\, k\lambda$.  Flux densities for calibrators
were determined by bootstrapping amplitude self-calibration solutions
to the absolute flux density calibrator.  Time-dependent amplitude
gain solutions were applied to \sgra.  Unfortunately, ALMA
observations did not include system temperature measurements on \sgra\
itself.  The pipeline software applied system temperatures obtained
for J1733-130 (NRAO 530) to \sgra, which introduces an error because of the
different atmospheric optical depth toward these two sources.  To
correct for this effect, we fitted system temperatures for all sources
in each epoch to determine the atmospheric optical depth and then
applied the correct system temperature to \sgra.  Typical changes to
the amplitude gain were a few percent, with a maximum value of 7\%.
Flux densities for \sgra\ were obtained through fitting a point-source of unknown flux to
visibilities on baselines longer than $50\, k\lambda$.  We report the
results in Table~\ref{tab:almaresults}.  We do not report statistical
errors on the flux densities because these are typically $\sim 1$ mJy,
much less than the likely uncertainty from gain calibration errors for
these bright sources.  We show ALMA spectra for \sgra\ and all sources
in Figure~\ref{fig:specalma}.

\begin{figure}[p!]
\includegraphics{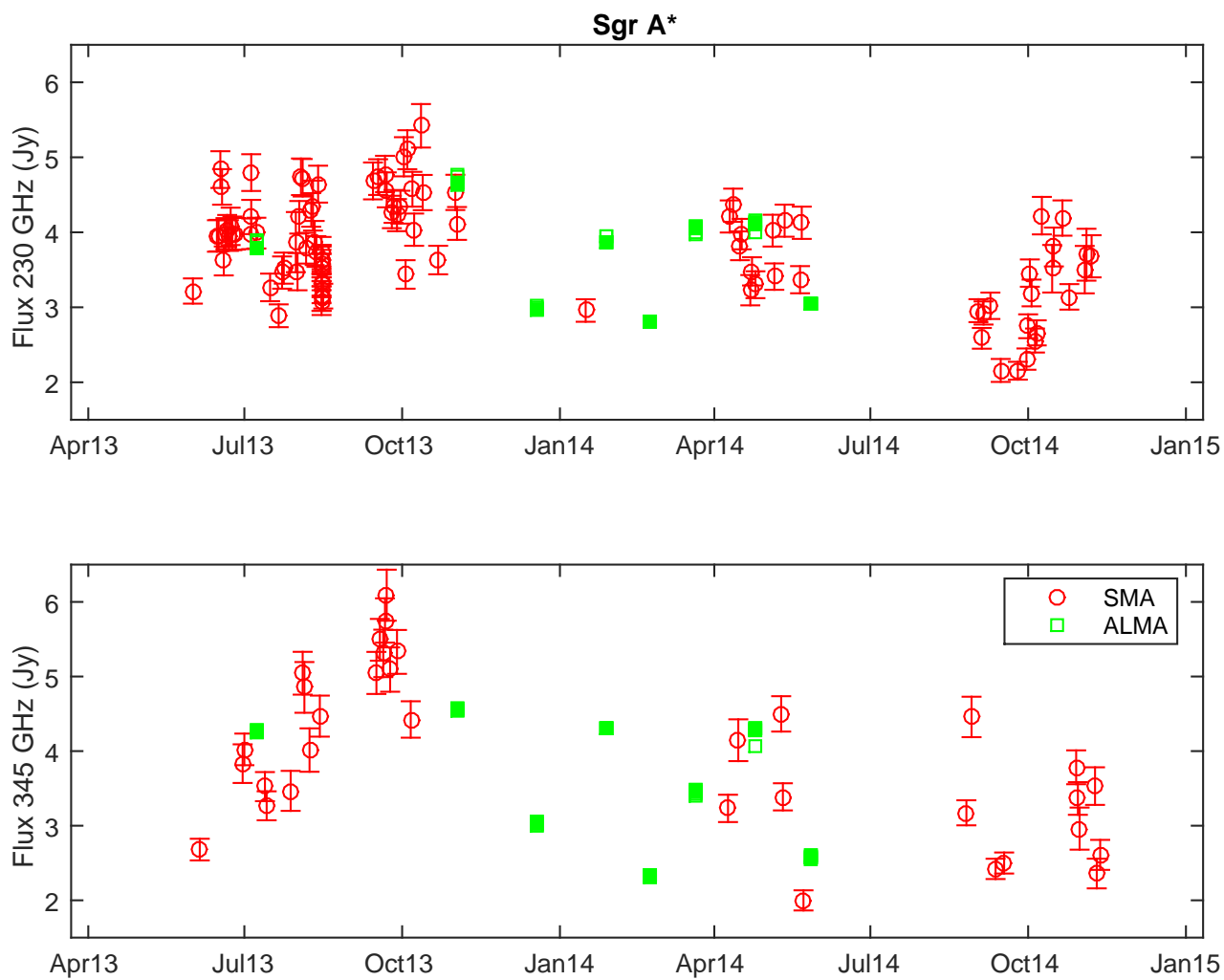}
\caption{
\label{fig:almasma}
ALMA and SMA light curves at 230 GHz (top) and 345 GHz (bottom) for \sgra.  Each ALMA spectral window is
plotted, covering 218 to 235 GHz (top) and 341 to 355 GHz (bottom).  The two lower frequency spectral windows
for each ALMA band are plotted with filled symbols.  These match most closely the plotted SMA flux densities,
which are in the range 212 to 241 GHz and 331 to 356 GHz, respectively, and have mean frequencies of 221 and 338 GHz.  
}
\end{figure}

\begin{figure}[p!]
\includegraphics{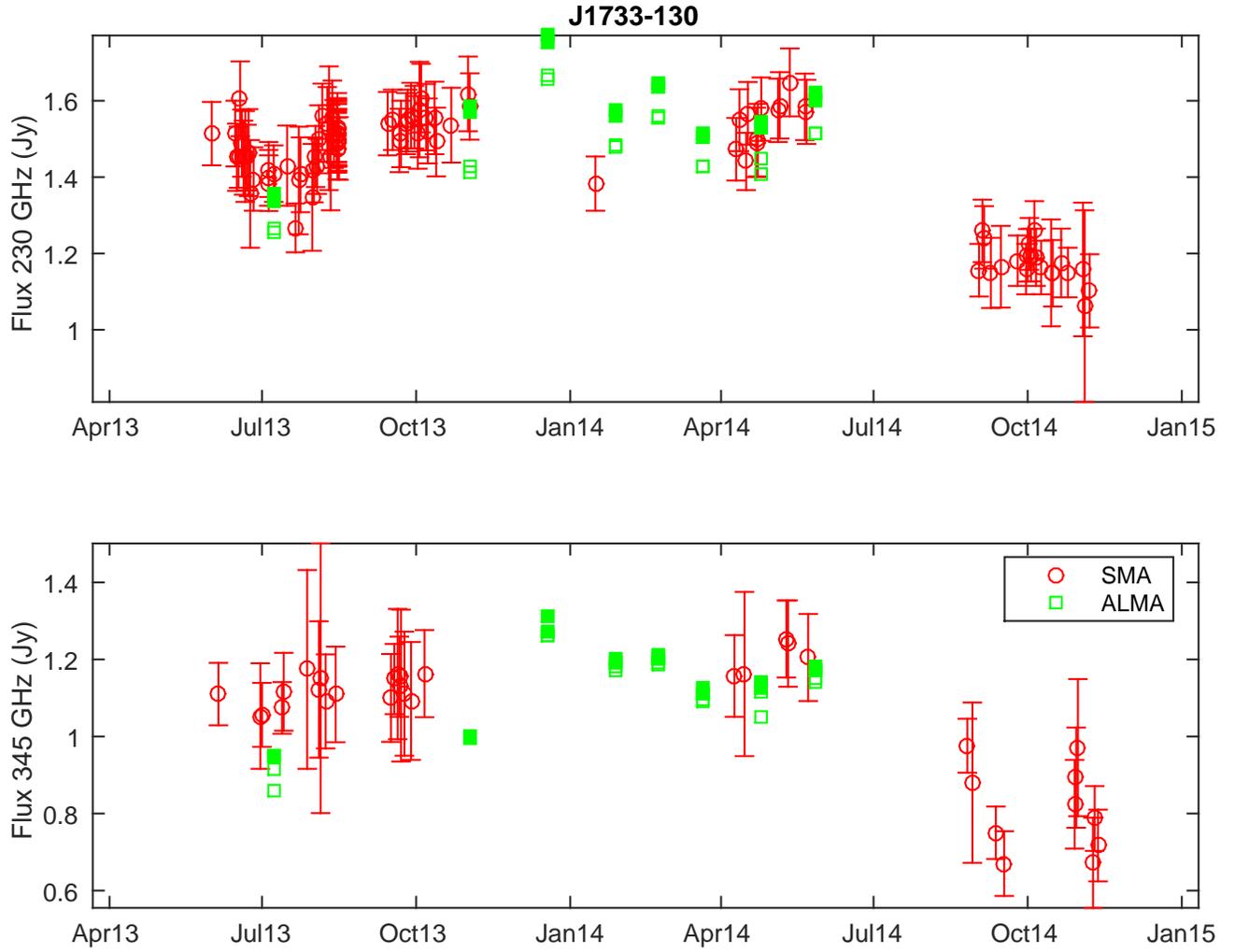}
\caption{
ALMA and SMA light curves at 230 GHz (top) and 345 GHz (bottom) for J1733-130.  
Symbols are the same as in Figure~\ref{fig:almasma}.
\label{fig:almasmaj1733}
}
\end{figure}

ALMA flux densities for the calibrators are consistent with recent SMA measurements
of flux densities.  We plot ALMA and SMA light curves for \sgra\ and J1733-130 in
Figures~\ref{fig:almasma} and \ref{fig:almasmaj1733}.  SMA data are described in the following
Section.  Where the data are contemporaneous, we see very good agreement in both
ALMA bands.


The variability of the calibrators sets an upper bound on the calibration accuracy
for \sgra.  J1733-130, J1700-261, and J1427-421 show 5\%, 9\%, and 14\% variability,
respectively, in the fitted intensity at 230 GHz assuming a power-law spectrum
for these sources.  Thus, we take 5\% as the systematic calibration error in the \sgra\
flux density.

\subsection{SMA Observations}

Observations of the flux density of \sgra\ were obtained using the
SMA, an interferometric array of eight 6-m diameter
radio dishes located near the summit of Mauna Kea, Hawaii \citep{2004ApJ...616L...1H}.  In most
cases the SMA was operating in a single receiver mode, providing 4 GHz
continuum bandwidth in a single polarization, in two sidebands, with
the array tuned to one of the three main SMA bands (1.3 mm, 1.1 mm and
870 micron).

Data were obtained during short ($\sim$20 min), self-contained observation
sets, in which observations of \sgra\ alternated with those of the
relatively nearby blazar J1733-130, and were bookended with
observations of a flux standard (typically Neptune).  
These observations were
``piggy backed'' on full-track scheduled observations-of-the-day for the SMA.
This explains the apparent randomness in sky frequencies. In addition,
the flux densities were estimated from a measurements with a single
linear polarization. The fractional linear polarization in Sgr A* is 
typically in the range of 5-10 percent and changes rapidly.  
The single polarization estimate of the flux density for a 
polarization of 10 percent has an rms error of 7 percent. This
error is comparable to the level of the statistical errors.
To this data set
were added more extensive observations over several hours in
dual polarization mode on 5 July and 15 Aug 2013, which allowed short-term
monitoring.

Data from each observing period were calibrated with the 
millimeter interferometer reduction (MIR) suite of
reduction routines developed by the 
SMA,\footnote{http://www.cfa.harvard.edu/rtdc/data/process}
involving removal of
visibility phase scatter due to atmospheric instability and scaling
visibility amplitude via comparison with the flux standard.  Neptune
was the primary flux standard used (85\% of observations), with Uranus
(13\%) and Titan (2\%) as the other standards.  For Uranus and Neptune,
we used the broadband spectral model from \citet{1993Icar..105..537G}.
However, due to the presence of broad CO absorption in the spectrum of
Neptune which is not included in the Griffin and Orton spectral fit,
we opted to use data that were at least 8 GHz separated from either the
CO(1-2) or CO(3-2) rotational transitions.  For Titan, we used the
spectral model developed by author Gurwell, which is now included in
the CASA data reduction package as well \citep[see][]{ButlerALMA}.  In all cases the expected
error on the flux density scale based upon using these models is ~5\%.
On the other hand, since we are most interested in tracking changes in
flux with time, the intrinsic bias in the flux density model is less
important.

The presence of gas along the line of sight toward \sgra\ leads to deep
absorption of the \sgra\ continuum at the CO transitions, as well as
their isotopologues.  For this reason, we masked spectral regions
around these transitions in the \sgra\ data.  Likewise, the structure
of the visibility data from \sgra\ indicated we were sensitive to broad
scale continuum emission from dust in the vicinity of \sgra.  In order
to limit our data to emission from \sgra\ itself, we used only 
visibility data from baselines that exceeded 35 k$\lambda$, or roughly
scales of 6\arcsec or finer, which was sufficient to isolate \sgra\ from the
broader diffuse emission.

Flux densities from the SMA for \sgra\ and J1733-130 are tabulated in
Table~\ref{tab:smaresults}.

\section{Results \label{sec:results}}

\begin{figure}[p!]
\includegraphics{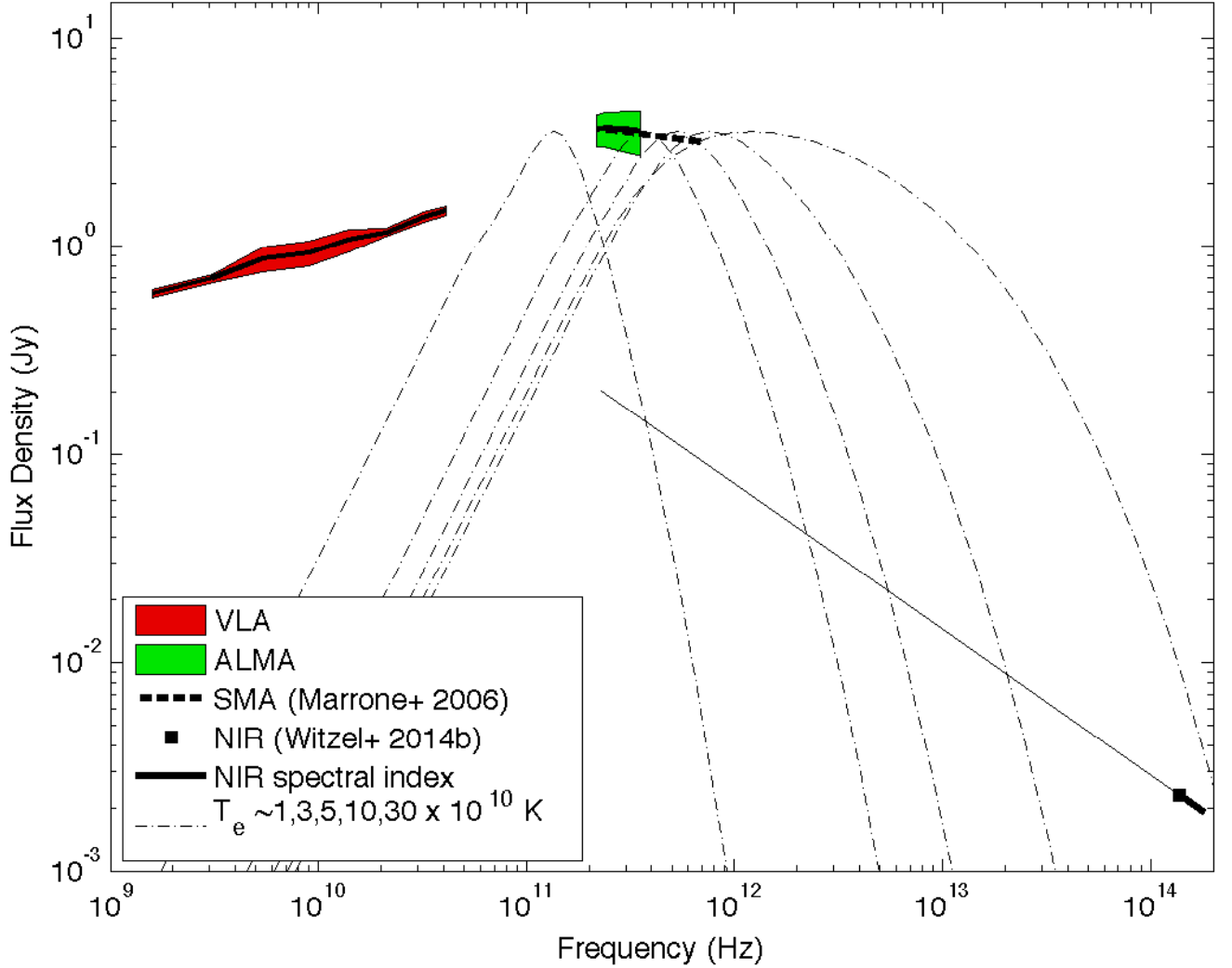}
\caption{ \label{fig:meanspec}
Mean spectrum of \sgra\ from VLA and ALMA observations.  The solid black line shows the mean
and the shaded regions show the $1\sigma$ boundaries for the flux
density based on the rms of the variations.  We also
plot a curve showing the 230 to 690 GHz spectrum as measured by \citet{2006JPhCS..54..354M}.  
We plot several model curves
representing synchrotron emission from relativistic thermal Maxwell-Juttner distributions.
We also plot NIR mean flux densities and spectral indices from \citet{2014IAUS..303..274W},  
along with a curve showing the extension of the NIR spectral index to shorter wavelengths (thin solid line).
See text for details.
}
\end{figure}

The mean spectrum of \sgra\ from VLA, ALMA, and SMA observations is presented in Table~\ref{tab:meanspec}
and Figure~\ref{fig:meanspec}, along with variability, minimum flux, and maximum flux.  
Variability is substantially stronger at millimeter wavelengths than at radio wavelengths.
The ratio of the rms variability to the total flux density, known as the modulation index, is 8\% at 40 GHz and below, while it is $\sim 20\%$ at millimeter
wavelengths.  This variability cannot be attributed to differences in calibration accuracy.

\begin{figure}[p!]
\includegraphics{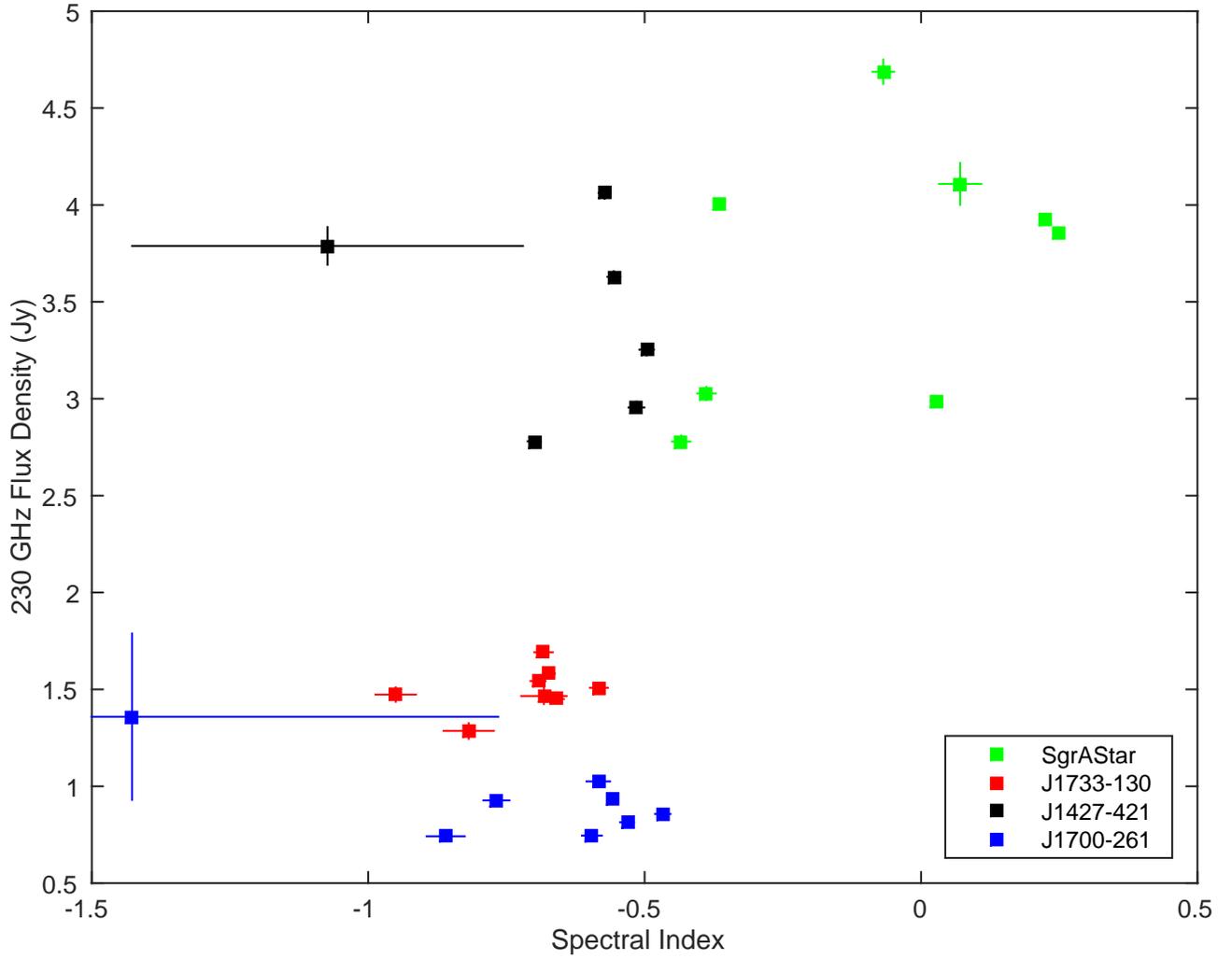}
\caption{
Fitted 230-GHz flux density and spectral index from the ALMA data (217 to 351 GHz) for \sgra\ and 
the calibrators.  
\label{fig:alphaalma}
}
\end{figure}

At VLA frequencies between 1 and 40 GHz, the mean spectral index $\alpha$ is $0.28 \pm 0.03$ ($S \propto \nu^\alpha$)  and shows evidence
for steepening at short wavelengths.  A mean spectral index of 0.5 between 40 and 218 GHz 
is required to connect the VLA and ALMA spectra, comparable to what has been
seen previously \citep{1998ApJ...499..731F}.
We also fit the ALMA data with a power-law spectrum
(Figure~\ref{fig:alphaalma}).  The mean spectral index for frequencies between 217 and 355 GHz is $\alpha = -0.06 \pm 0.26$, which is 
flatter and more variable than those of the calibrators.  The calibrators
have mean spectral indices between 217 and 355 GHz of $-0.67 \pm 0.06$, $-0.56 \pm 0.08$, and $-0.57 \pm 0.07$
for J1733-130, J1700-261, and J1427-421, respectively.

\citet{2006JPhCS..54..354M} found that the 230 to 690 GHz spectral index ranged in four epochs from -0.4 to +0.2, with
a mean of -0.13.  Our results span the same range of $\alpha$ and are statistically indistinguishably
from those of \citet{2006JPhCS..54..354M}; that is, the spectrum is consistent with 
a flat or slightly declining power-law spectrum above 230 GHz.  Assuming stationary statistics for 
\sgra\ between the early observations and these new observations, we conclude that there is no evidence
for spectral curvature or greater steepening in the spectrum between 230 and 690 GHz.

The millimeter and submillimeter spectrum of \sgra\ indicates that the
optical depth is $\sim 1$ at frequencies as high as 690 GHz.  The result suggests that the source
must be composed of stratified regions near the optically
thick-to-thin transition in order to produce a near-flat synchrotron
spectrum over this broad frequency range.  A single optically thin 
synchrotron component cannot reproduce the spectrum.  This spectrum can be
explained in the context of the classical 
\citet{BlandfordKoenigl1979} jet model, but also by inflowing
accretion flow models with radially evolving density and magnetic
fields together with some non-thermal particles 
\citep{YuanQuataertNarayan2003}.

Despite this degeneracy in the emission geometry, it is clear that
above 350 GHz, Sgr A* has to become optically thin.   This
conclusion is supported by the transition to higher linear polarization
at higher frequencies
\citep{Boweretal2003,2006JPhCS..54..354M}, and the
increased variability and power-law spectrum seen in the infrared band
\citep[e.g.,][]{2003Natur.425..934G,2004ApJ...601L.159G,2006A&A...450..535E,2012ApJS..203...18W,2014IAUS..303..274W}.

Our simultaneous radio through submillimeter data can provide some
interesting constraints on the underlying particle distribution, when
put into context with the infrared measurements.  In
Fig.~\ref{fig:meanspec} we include the average IR flux based on VLT
and Keck measurements \citep{2012ApJS..203...18W,2014IAUS..303..274W}.
These results are consistent with a mean flux density of $\sim 3$ mJy
in the $K_s$ band and a stable spectral index $\alpha\sim -0.6$
between $K_s$ and $H$ band.  Previous estimates of the NIR spectral
index appear to show steeper spectral indices but may have suffered
from non-simultaneous observations of the variable flux density
\citep[e.g.,][]{2011A&A...532A..26B}.  To illustrate the new limits on
the radiating particle distributions, we plot for reference a series
of synchrotron spectra from mildly relativistic thermal electrons in
equipartition with the magnetic field, ranging from
$1-30\times10^{10}$ K, assuming an emission region of size $3r_g$.
The observed brightness temperature of Sgr A* at 230 GHz is $\sim 6 \times 10^{10}$ K
\citep{2008Natur.455...78D}.
These are fiducial values spanning the range from the literature
\citep[see, e.g.,][and references therein]{2013CQGra..30x4003F}, to
illustrate how stratified, self-absorbed regions can add to a
near-flat spectrum, but also to place some limits on the fraction and
distribution of non-thermal particles present.  The peak of a given
temperature component can shift by a factor of a few in frequency for
different emission size regions or energy partition between radiating
particles and the magnetic energy density.  However, it is clear that a peak
temperature (in the plasma closest to the black hole) below a
few $10^{10}$ K cannot account for the extension of the
self-absorbed spectrum through the ALMA range shown, and that a peak
temperature much higher than a few $10^{11}$ K would violate the
IR limits.  It is also clear from these spectra that a purely thermal
distribution cannot account for the IR flux.

The intersection of the IR power-law with the intermediate ranges
indicates that a non-thermal ``tail'' of particles is present, with
normalization on the order of $\le1$\% of the thermal peak.  This
fraction is consistent with the results of fits to Sgr A* in
quiescence, using a self-consistent calculation of the particle
distribution given a mechanism for injecting a power-law of nonthermal
particles \citep{Dibietal2014} .  Recent work exploring second-order
acceleration processes in turbulent, magnetized plasmas by
\cite{Lynnetal2014} shows the self-consistent production of a such a
nonthermal tail; however, its slope and relation to the thermal peak
may be difficult to match to these newest constraints.  Therefore the
combination of simultaneous ALMA (eventually using even higher frequency bands)
and IR data offers the best constraints yet on the shape of the
radiating particle distributions and nonthermal fraction.  The results
presented here can already be used to guide implementation of the
inclusion of particle acceleration in semi-analytical models
\citep[e.g.][]{2001A&A...379L..13M,2003ApJ...598..301Y,BroderickLoeb2009}
and in the so-called ``painting'' of GRMHD simulations to produce
images
\citep[e.g.,][]{2012MNRAS.421.1517D,MoscibrodzkaFalcke2013,Moscibrodzkaetal2014,Chanetal2014}.
Simultaneous observations between ALMA, IR, and X-ray, particularly
during flaring, would provide strong constraints on the upper extreme
of the nonthermal population and accelerating mechanisms.

\subsection{Presence of a Bow Shock}

Bow shock emission from the interaction of G2 with the accretion flow
was anticipated to appear and peak months before the center of mass
reached periastron \citep{2012ApJ...757L..20N}.  The bow shock would
excite nonthermal electrons that then produce synchrotron radiation in
the magnetic field of the accretion flow.  The synchrotron radiation
is expected to peak at frequencies near 1 GHz and decline with an
optically thin power-law index $\sim -0.7$.  The predicted peak flux
density of the radio light curve scales with a number of factors,
including the relative velocity of the cloud, the accretion flow
density profile, the efficiency of nonthermal electron energy
production, and the timescale for electrons to cool.  Perhaps the most
important term is the cross-sectional area of the shock as it impacts
the accretion flow.  \citet{2013MNRAS.433.2165S} use the geometrical
size of G2 as observed at large radii, $\sim 3 \times 10^{30} {\rm\,
  cm^2}$.  \citet{2014ApJ...783...31S}, on the other hand, constructs
a magnetically arrested, tidally distorted cloud, which has an area
$\sim 10^{29} {\rm\, cm^2}$.  \citet{2013MNRAS.436.1955C} model G2 as
a wind driven from a hidden star with a radius that shrinks as the
external pressure in the accretion flow increases, leading to a
minimum area at periastron of $\sim 10^{28} {\rm\, cm^2}$.  The flux
density in all models scales linearly with the area.  Thus,
predictions of the 1 GHz excess flux density excess range from $>10$ Jy for some 
models to $<0.1$ Jy for others.

Synchrotron lifetimes for the radiating electrons in the accretion
inflow are long, so the timescale for the radio emission is determined
by the dynamics of the bow shock.  \citet{2014ApJ...783...31S}
predicts a characteristic timescale of four months, which is
comparable to predictions in other models.

\begin{figure}[p!]
\includegraphics{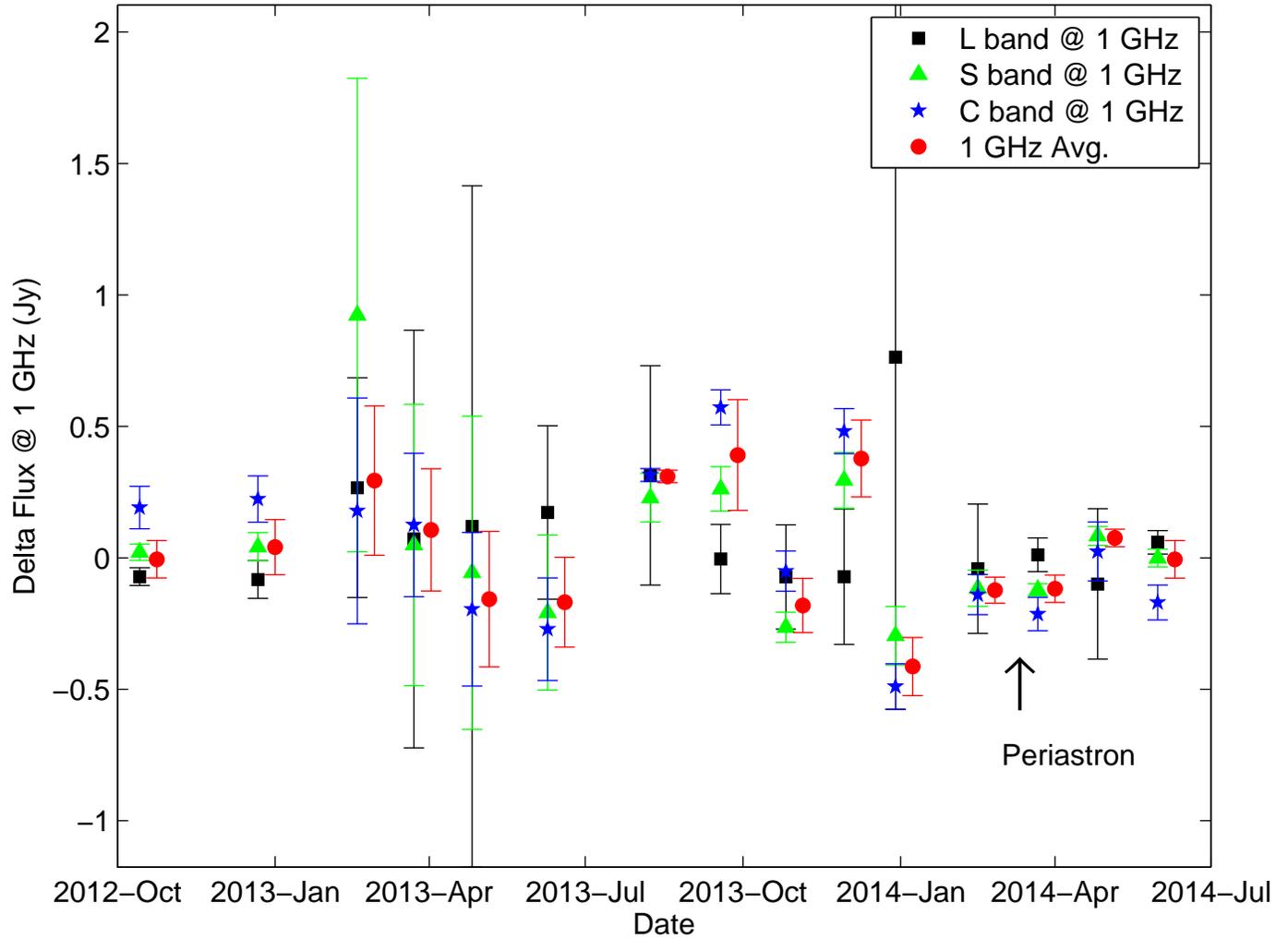}
\caption{
Estimates of the 1.0 GHz flux density excess above the average flux density.  We extrapolate
L, S, and C band (1.5, 3.1, 5.4 GHz) flux density excesses to 1 GHz using a power-law index of -0.7.  The
red symbols show the average of all three bands.  Time is offset for the averages by 10 days for clarity
in the figure.  The arrow marks the estimated time of
periastron \citep{2015ApJ...798..111P}.
\label{fig:bowshock}
}
\end{figure}

The L band data alone in the first and last epochs of our experiment
show that no significant secular change has occurred over the course of this
experiment.  Between October 2012 and May 2014, the 1.5 GHz flux density
changed by at most 12 mJy (2\%).  We can examine in more detail whether there is evidence
for shorter time scale variations through an analysis of all the measurements
in the three lowest VLA frequency bands.
We compute estimates of the flux density excess above the average at 1.0 GHz using 
low frequency measurements (Figure~\ref{fig:bowshock}).  We extrapolate the flux density
excess at 1.5, 3.1, and 5.4 GHz
to 1 GHz, assuming a power-law spectrum with index $\alpha=-0.7$, appropriate for
optically thin synchrotron emission. At frequencies above 5 GHz, 
intrinsic variability rises rapidly and is likely to exceed estimates of any bow shock-related
variability.  1.5 GHz results alone are sometimes only weakly sensitive to variations because of 
limited overlap in $(u,v)$ coverage for the most compact configurations.  

The average 1 GHz flux density excess  computed in this way is only
significantly non-zero for a single epoch, 2013 August 08, with a
value of $0.310 \pm 0.024$ Jy.  The excess is positive and comparable
for all three bands and very high significance at 5 GHz.  The
subsequent epoch, 2013 September 18, has a comparable value, $0.391
\pm 0.211$ Jy, but much lower significance.  The total flux density
spectrum for 2013 August 08 is one of the most elevated across the
band, with a peak flux density of 1.8 Jy at 40 GHz.  These
observations were obtained in a compact configuration and no total
flux density measurements were obtained at 1.4 and 3.1 GHz.

For 11 of 15 epochs, the 95\% confidence upper limit on 1 GHz flux density excess is 0.4 Jy.  
For the remaining four epochs, the upper limit is 0.8 Jy.
A four-month moving average of the average 1 GHz flux has a peak at 0.3 Jy, centered
on the 2013 August 08 epoch, and typical values of 0.1 Jy.
These results clearly exclude an object with size $3 \times 10^{30} {\rm\, cm^2}$.  
At 95\% confidence, we can set an upper limit to the size of G2 of $\sim 2\times 10^{29} {\rm\, cm^2}$.  
This can be attributed to the magnetically arrested cloud derived by \citet{2014ApJ...783...31S}
or due to a pressure-confined stellar wind \citep{2013MNRAS.436.1955C}.
Alternatively, the detailed model of the accretion flow structure could be in error.

\subsection{Changes in the Mean Spectrum}

The new data appear to show higher radio flux densities at frequencies $> 2$ GHz over historical averages
(Figure~\ref{fig:specvla} and Table~\ref{tab:meanspec}).  The 40.9 GHz flux density, for instance,
has a mean of 1.5 Jy and a range of 1.3 to 1.9 Jy, whereas \citet{2004AJ....127.3399H} found a 43 GHz flux density
mean of 1.0 Jy and a range of 0.6 to 1.9 Jy. Similar results with smaller amplitudes differences are 
apparent at lower frequencies.  The flux densities from \citet{1998ApJ...499..731F} and \citet{2001ApJ...547L..29Z} 
are higher than those of \citet{2004AJ....127.3399H} but still show a lower average flux density at most frequencies
relative to the new VLA data.  \citet{2001ApJ...547L..29Z} presented many epochs of archival data obtained between
1979 and 1999.
\citet{2004AJ....127.3399H} reported over 100 epochs with regular monitoring from 2000 to 2003, 
while \citet{1998ApJ...499..731F} is from a single-epoch simultaneous spectrum.  

Over the whole radio spectrum, we infer an increase of $\sim 30\%$
relative to \citet{2004AJ....127.3399H} and $\sim 10\%$ relative to
\citet{2001ApJ...547L..29Z} in the total flux density.  These
differences are comparable to the variability between
\citet{2004AJ....127.3399H} and \citet{2001ApJ...547L..29Z}  and
consistent with variability seen in low-luminosity AGN (LLAGN) in general
\citep{2002A&A...392...53N,2008ARA&A..46..475H}. Thus, these differences cannot be attributed to any effects associated
with the G2 periastron.
The mean flux density that we measure at 21 GHz, $1.16 \pm 0.05$ Jy, is comparable to the 22 GHz flux
density measured over a similar time period with the Japanese VLBI Network, $1.23 \pm 0.33$ Jy
\citep{2015ApJ...798L...6T}.  The lower variability in our result may reflect some of the difficulties
of providing accurate amplitude calibration for VLBI arrays.

A comparison against historical millimeter and submillimeter flux densities also does not
reveal any significant change.  \citet{2014MNRAS.442.2797D} presents historical data for \sgra\ at
230, 345, and 690 GHz with dates from 2001 through 2012.  The flux densities from this paper
span timescales from minutes to years and were obtained with multiple telescopes
with varying calibration accuracy.  The mean ALMA Band 6 (230 GHz) flux density is $3.6 \pm 0.6$ Jy while
the mean historical flux density at 230 GHz is $3.4 \pm 0.5$ Jy.  Similarly, for Band 7 (345 GHz) and historical
345 GHz data,
we find $3.6 \pm 0.8$ Jy and $3.0 \pm 0.6$ Jy, respectively.  Results from the SMA
data are comparable.  Thus, increases in the millimeter and submillimeter
mean flux density are no more than 20\% relative to historical
averages, again consistent with typical LLAGN variability levels.  

A Kolmogorov-Smirnov test shows that the ALMA and SMA measurements are
consistent ($p \sim 0.1$) with originating from the same
distribution.  The ALMA and SMA measurements differ from the
\citet{2014MNRAS.442.2797D} measurements with varying significance.
SMA 230 and 345 GHz data differ from the \citet{2014MNRAS.442.2797D}
distributions with $p < 10^{-2}$, while the Band 6 (230 GHz) and Dexter 230 GHz
and Band 7 and 345 GHz distributions differ with significances of
$p=0.03$ and $p<10^{-3}$.  These variations may be partly the result
of different calibration accuracies and as the result of evolution of
the light curves over $\sim 10$ yr.  Similar to the case with the
radio spectrum, there is no clear evidence that the millimeter and submillimeter
changes are the result of the G2 periastron.

We have analyzed the light curve data for our measurements near 220 GHz from both
ALMA and the SMA. There are 104 data points spanning an interval of 524 
days with the shortest interval being 0.02 days. The data were binned
in a sparsely filled array of 32,768 elements and Fourier transformed.  
The resulting spectrum was averaged into logarithmically 
spaced frequency intervals.  The spectrum has a simple
power law form characterized as $S \propto f^\alpha$  over a frequency 
interval of 0.04 to 2 cycle/day. $\alpha$  was found to be $-0.15 \pm 0.08$.
The error bar was determined by bootstrap resampling.  This result is
intuitively obvious from the light curve, i.e. the variations over a period
of a few day are uncorrelated and have the same range as the variations 
over the entire data span. The  spectral variations from historical 
data at millimeter/submillimeter wavelengths are known to approximate 
white noise at frequencies less than 3 cycles/day, with a steepening power 
law above that frequency \citep{2014MNRAS.442.2797D}.  It is plausible that 
an enhancement of emission caused by G2 might have 
produced long term variations indicative of red noise. 
However, there is no evidence for this effect.

\section{Discussion and Conclusions \label{sec:discussion}}

The spectrum of \sgra\ from radio to submillimeter wavelengths has
remained remarkably stable over the past 30 years.  Our new VLA, ALMA,
and SMA results demonstrate that this stability has continued
throughout the apparent periastron passage of G2.  In the case of low-frequency
emission, the absence of significant change permits us to constrain
the size of the ensuing bow shock to be a factor of 30 smaller than the
apparent G2 size observed with NIR wavelengths at larger distances
from the black hole.  These observations do not resolve the nature of
G2.  The smaller bow shock size is consistent with both cloud and
stellar-wind models.  

\citet{2014ApJ...796L...8W} propose a model
that may reconcile the apparently discrepant NIR continuum and spectroscopic
observations and may also be consistent with our results.  $L^\prime$
continuum observations reveal a compact source while Br-$\gamma$
observations
reveal an extended source that appears to be tidally disrupted.
\citet{2014ApJ...796L...8W} argue that the continuum emission is the
photosphere of a binary stellar merger, while the ionized emission
represents a smaller tidally disrupted tail.  The $L^\prime$ source has
a cross-sectional area of $\sim 6 \times 10^{27}$ cm$^{2}$, much less
than our upper limit of $2 \times 10^{29}$ cm$^{2}$.  
However, the size of the extended Br-$\gamma$ emission, which is consistent
between VLT and Keck observations, 
primarily determines the creation of a bow shock.
Some form of confinement
for the diffuse gas appears to be important for attenuating the shock
amplitude.

The lack of any enhanced short wavelength radio and millimeter emission
from the black hole is also consistent with a compact size for G2.  Simulations
have shown that more extended objects are more readily disrupted and would therefore be more likely to have some gas
initiate enhanced accretion onto the black hole.  

At shorter wavelengths, the mean flux density has not increased by
more than $\sim 20\%$ relative to historical averages.  This is within
the range of variations seen in the past  and in other LLAGN.  Thus,
we cannot attribute any increase in the flux density to a G2-induced
enhancement of the accretion rate.  The amplitude and timescale for a
change in the accretion rate, should G2 be fully disrupted, are
uncertain.  In order for the disrupted material to enhance the flux
density, the gas must reach a radius of a few Schwarzschild radii
to contribute to the accretion flow and/or enhance jet power.
The free-fall time from $1000 R_{Sch}$ is $t_{ff} \sim 0.1$ yr.  
If a fraction $f$
of the material were on a plunging orbit due to tidal disruption or a
spread in orbital characteristics, we would have expected an
enhancement of the overall flux density during these observations.
The flux density should scale as
\begin{equation}
{\Delta S \over S } = \left( {fM_{G2} t_{ff}^{-1} \over \dot{M}} \right)^\gamma,
\end{equation}
where $\gamma\sim 1$ is a model-dependent constant 
for jet and accretion disk models \citep{1993A&A...278L...1F,2007MNRAS.379.1519M,2012ApJ...752L...1M}.  For a G2
mass of $M_{G2} \approx 3 M_\earth = 10^{-5} M_\odot$ 
appropriate for the gas-cloud model
and a steady-state
accretion rate onto \sgra\ of $\dot{M}\sim 10^{-8} M_\odot\,{\rm
  yr^{-1}}$, and ${\Delta S \over S }\lsim 0.2$, we find $f \lsim 2
\times 10^{-5}$ for $\gamma=1$.  That is, very little of G2's mass
has yet accreted directly onto the black hole.  Longer term monitoring
(10 yr) will be sensitive to $f \gsim 10^{-3}$ reaching the black hole
through an increase of the average flux density.  
The bulk of the material is more likely to accrete on the viscous time
scale, $t_v >> t_{ff}$.
$t_v \sim t_{ff} \alpha_{v}^{-1} (H/R)^{-2}$, where $\alpha_{v}\sim 0.1$ -- 1
is the viscous parameter
and $H$ is the height at radius $R$.  For reasonable parameters, $t_v \sim 0.1$ -- 100 yr.
Continued monitoring of the radio through millimeter flux densities can constrain further the
properties and origin of G2 as well as the properties of the accretion flow
at large radii.

The long-term stability of the radio/millimeter spectrum indicates that episodic events such as a fully disrupted cloud
falling onto the black hole are rare.  Such cloud disruptions may take place but they must have timescales that are long
compared to the free-fall time.

We confirm a flat spectrum for \sgra\ emission in the millimeter/submillimeter regime.
The emission is likely to originate from a stratified region with optical depth in
transition between optically thick and thin.
Optically thick emission at frequencies of $>230$ GHz may influence images obtained with
the Event Horizon Telescope \citep{2013arXiv1309.3519F}.    Accretion disk models with tilted disks and/or high ratios of ion to electron temperature
produce larger regions with optical depth greater than 1 \citep{2009ApJ...706..497M,2013MNRAS.432.2252D}.
Higher optical depths may obscure or complicate the interpretation of 
gravitational lensing effects such as the black hole shadow.  On the other hand,
higher optical depth in regions that are offset from the black hole as expected from
Doppler boosting may enhance the contrast associated with geometric features.
New ALMA measurements of the spectrum at frequencies from 100 to 950 GHz can resolve the 
uncertainty in the spectrum between the submillimeter and the NIR regime and assist in 
optimization of EHT observations.

\acknowledgements
The National Radio Astronomy Observatory is a facility of the National Science Foundation operated under cooperative agreement by Associated Universities, Inc.
This paper makes use of the following ALMA data: ADS/JAO.ALMA\# 2012.1.00635.S. ALMA is a partnership of ESO (representing its member states), NSF (USA) and NINS (Japan), together with NRC (Canada) and NSC and ASIAA (Taiwan), in cooperation with the Republic of Chile. The Joint ALMA Observatory is operated by ESO, AUI/NRAO and NAOJ.
The Submillimeter Array is a joint project between the Smithsonian Astrophysical Observatory and the Academia Sinica Institute of Astronomy and Astrophysics and is funded by the Smithsonian Institution and the Academia Sinica.
We thank Michael Johnson for  helpful discussions.


\begin{deluxetable}{lrrrr}
\tablecaption{VLA Observations \label{tab:vlaobs}}
\tablehead{
\colhead{Epoch} & \colhead{UT} & \colhead{Beam Q band} & \colhead{Beam L band}  & Configuration\\
                &              & (arcsec$^2$, deg) & (arcsec$^2$, deg) \\ 
}
\startdata
20121014 & 23:16 - 00:25 & $  0.13 \times   0.11 $ ,   11.7 & $  2.97 \times  2.45 $,   45.1 & A \\
20121222 & 17:30 - 18:39 & $  0.17 \times   0.06 $ ,  -15.8 & $  2.74 \times  1.45 $,  -12.5 & A \\
20130218 & 13:42 - 14:51 & $  4.36 \times   1.65 $ ,  -13.6 & $ 79.21 \times 74.20 $,  -72.9 & D \\
20130323 & 12:02 - 13:11 & $  4.39 \times   2.77 $ ,   29.3 & $ 90.13 \times 52.88 $,   36.9 & D \\
20130426 & 10:56 - 12:05 & $  4.43 \times   2.83 $ ,   28.0 & $ 88.42 \times 46.85 $,   50.2 & D \\
20130609 & 06:25 - 07:34 & $  1.53 \times   1.12 $ ,   42.9 & $ 103.35 \times 25.77 $,  -26.2 & C \\
20130808 & 03:30 - 04:39 & $  1.39 \times   0.93 $ ,   24.2 & $ 38.22 \times 19.51 $,   29.1 & C \\
20130918 & 01:33 - 02:42 & $  0.75 \times   0.40 $ ,   63.1 & $ 26.32 \times  8.73 $,   32.4 & CnB \\
20131026 & 23:14 - 00:24 & $  0.43 \times   0.27 $ ,   28.9 & $ 12.05 \times  5.85 $,   23.4 & B \\
20131129 & 19:46 - 20:55 & $  0.41 \times   0.28 $ ,   23.4 & $  9.05 \times  5.87 $,   26.8 & B \\
20131229 & 17:48 - 18:57 & $  0.43 \times   0.30 $ ,   33.7 & $ 38.08 \times 14.34 $,   -1.3 & B \\
20140215 & 14:11 - 15:21 & $  0.14 \times   0.10 $ ,   37.3 & $  3.37 \times  3.00 $, -101.8 & BnA \\
20140322 & 13:26 - 14:35 & $  0.13 \times   0.09 $ ,   22.0 & $  3.26 \times  1.53 $,   24.0 & A \\
20140426 & 10:19 - 11:28 & $  0.12 \times   0.08 $ ,   27.0 & $  3.01 \times  2.11 $,  -27.4 & A \\
20140531 & 08:11 - 09:20 & $  0.28 \times   0.15 $ ,   43.6 & $  3.41 \times  2.07 $,  -28.1 & A \\
\enddata
\end{deluxetable}

\begin{deluxetable}{llrrr}
\tablecaption{Frequency Coverage \label{tab:frequencytable}}
\tablehead{
\colhead{Tel.} & \colhead{Band} & \colhead{$\nu_{lower}$} & \colhead{$\nu_{upper}$} & \colhead{$N_{ch}$}  \\
               &                & \colhead{(GHz)} & \colhead{(GHz)} \\
}
\startdata
VLA & L & 994.0 & 2006.0  & 1024 \\
VLA & S & 1988.0 & 3948.0  & 1024 \\
VLA & C & 4488.0 & 6448.0  & 1024 \\
VLA & X & 7988.0 & 9948.0  & 1024 \\
VLA & U & 12988.0 & 14948.0  & 1024 \\
VLA & K & 20188.0 & 22148.0  & 1024 \\
VLA & A & 32008.0 & 31968.0  & 1024 \\
VLA & Q & 39988.0 & 41948.0  & 1024 \\
ALMA & B6-1 & 217.0 & 219.0 & 128 \\
ALMA & B6-2 & 219.0 & 221.0 & 128    \\
ALMA & B6-3 & 231.0 & 232.9 & 3840\\
ALMA & B6-4 & 232.8 & 234.8 & 128    \\
ALMA & B7-1 & 340.6 & 342.6 & 128 \\
ALMA & B7-2 & 342.6 & 344.6 & 128 \\
ALMA & B7-3 & 350.7 & 352.7 & 128 \\
ALMA & B7-4 & 352.7 & 354.6 & 3840\\
\enddata
\end{deluxetable}

\begin{deluxetable}{lrrr}
\tablecaption{VLA Results \label{tab:vlaresults}}
\tablehead{
\colhead{Epoch} & \colhead{Frequency} & \colhead{Flux Density} & \colhead{Delta Flux Density} \\
	        & \colhead{(GHz)} & \colhead{(Jy)} & \colhead{(Jy)}
}
\startdata
20121014 &  1.5 & $ 0.578 \pm 0.016 $ & $ -0.055 \pm 0.026 $ \\ 
20121014 &  3.0 & $ 0.739 \pm 0.045 $ & $  0.010 \pm 0.014 $ \\ 
20121014 &  5.4 & $ 0.867 \pm 0.031 $ & $  0.059 \pm 0.025 $ \\ 
20121014 &  8.9 & $ 0.982 \pm 0.031 $ & $  0.093 \pm 0.005 $ \\ 
20121014 & 13.9 & $ 1.092 \pm 0.006 $ & $  0.078 \pm 0.007 $ \\ 
20121014 & 21.1 & $ 1.213 \pm 0.007 $ & $  0.045 \pm 0.002 $ \\ 
20121014 & 32.0 & $ 1.370 \pm 0.011 $ & $ -0.041 \pm 0.009 $ \\ 
20121014 & 40.9 & $ 1.427 \pm 0.004 $ & $ -0.110 \pm 0.008 $ \\ 
20121222 &  1.5 & $ 0.558 \pm 0.029 $ & $ -0.062 \pm 0.054 $ \\ 
20121222 &  3.0 & $ 0.691 \pm 0.008 $ & $  0.020 \pm 0.025 $ \\ 
20121222 &  5.4 & $ 0.876 \pm 0.029 $ & $  0.068 \pm 0.027 $ \\ 
20121222 &  8.9 & $ 0.952 \pm 0.037 $ & $  0.037 \pm 0.015 $ \\ 
20121222 & 13.9 & $ 1.009 \pm 0.007 $ & $ -0.004 \pm 0.011 $ \\ 
20121222 & 21.1 & $ 1.170 \pm 0.012 $ & $  0.010 \pm 0.015 $ \\ 
20121222 & 32.0 & $ 1.525 \pm 0.102 $ & $  0.022 \pm 0.020 $ \\ 
20121222 & 40.9 & $ 1.478 \pm 0.052 $ & $  0.043 \pm 0.043 $ \\ 
20130218 &  1.5 & \dots               & $  0.202 \pm 0.317 $ \\ 
20130218 &  3.0 & \dots               & $  0.424 \pm 0.413 $ \\ 
20130218 &  5.4 & \dots               & $  0.055 \pm 0.131 $ \\ 
20130218 &  8.9 & \dots               & $  0.081 \pm 0.075 $ \\ 
20130218 & 13.9 & $ 1.749 \pm 0.088 $ & $  0.186 \pm 0.028 $ \\ 
20130218 & 21.1 & $ 1.421 \pm 0.021 $ & $  0.140 \pm 0.011 $ \\ 
20130218 & 32.0 & $ 1.505 \pm 0.015 $ & $  0.059 \pm 0.008 $ \\ 
20130218 & 40.9 & $ 1.666 \pm 0.040 $ & $  0.163 \pm 0.015 $ \\ 
\tablebreak
20130323 &  1.5 & \dots               & $  0.055 \pm 0.610 $ \\ 
20130323 &  3.0 & \dots               & $  0.023 \pm 0.246 $ \\ 
20130323 &  5.4 & \dots               & $  0.038 \pm 0.083 $ \\ 
20130323 &  8.9 & \dots               & $ -0.020 \pm 0.026 $ \\ 
20130323 & 13.9 & $ 1.455 \pm 0.013 $ & $ -0.111 \pm 0.021 $ \\ 
20130323 & 21.1 & $ 1.344 \pm 0.035 $ & $ -0.134 \pm 0.015 $ \\ 
20130323 & 32.0 & $ 1.274 \pm 0.017 $ & $ -0.237 \pm 0.012 $ \\ 
20130323 & 40.9 & $ 1.395 \pm 0.018 $ & $ -0.211 \pm 0.009 $ \\ 
20130426 &  1.5 & \dots               & $  0.090 \pm 0.980 $ \\ 
20130426 &  3.0 & \dots               & $ -0.026 \pm 0.273 $ \\ 
20130426 &  5.4 & \dots               & $ -0.060 \pm 0.090 $ \\ 
20130426 &  8.9 & \dots               & $  0.041 \pm 0.037 $ \\ 
20130426 & 13.9 & \dots               & $ -0.009 \pm 0.021 $ \\ 
20130426 & 21.1 & $ 1.562 \pm 0.029 $ & $  0.122 \pm 0.009 $ \\ 
20130426 & 32.0 & $ 1.648 \pm 0.017 $ & $  0.254 \pm 0.015 $ \\ 
20130426 & 40.9 & $ 1.882 \pm 0.132 $ & $  0.374 \pm 0.023 $ \\ 
20130609 &  1.5 & \dots               & $  0.131 \pm 0.250 $ \\ 
20130609 &  3.1 & \dots               & $ -0.094 \pm 0.133 $ \\ 
20130609 &  5.4 & $ 1.097 \pm 0.105 $ & $ -0.083 \pm 0.060 $ \\ 
20130609 &  8.9 & $ 1.059 \pm 0.069 $ & $ -0.070 \pm 0.053 $ \\ 
20130609 & 13.9 & $ 1.137 \pm 0.040 $ & $ -0.060 \pm 0.006 $ \\ 
20130609 & 21.1 & $ 1.037 \pm 0.219 $ & $ -0.183 \pm 0.045 $ \\ 
20130609 & 32.0 & $ 1.286 \pm 0.051 $ & $ -0.108 \pm 0.052 $ \\ 
20130609 & 40.9 & $ 1.278 \pm 0.058 $ & $ -0.249 \pm 0.059 $ \\ 
\tablebreak
20130808 &  1.5 & \dots               & $  0.240 \pm 0.319 $ \\ 
20130808 &  3.0 & \dots               & $  0.106 \pm 0.043 $ \\ 
20130808 &  5.4 & $ 0.835 \pm 0.086 $ & $  0.096 \pm 0.007 $ \\ 
20130808 &  9.4 & $ 1.276 \pm 0.022 $ & $  0.055 \pm 0.008 $ \\ 
20130808 & 13.9 & $ 1.273 \pm 0.038 $ & $  0.118 \pm 0.013 $ \\ 
20130808 & 21.1 & $ 1.335 \pm 0.031 $ & $  0.161 \pm 0.013 $ \\ 
20130808 & 32.0 & $ 1.596 \pm 0.006 $ & $  0.225 \pm 0.010 $ \\ 
20130808 & 40.9 & $ 1.826 \pm 0.014 $ & $  0.391 \pm 0.011 $ \\ 
20130918 &  1.5 & \dots               & $ -0.003 \pm 0.100 $ \\ 
20130918 &  3.0 & $ 0.796 \pm 0.025 $ & $  0.120 \pm 0.039 $ \\ 
20130918 &  5.4 & $ 1.134 \pm 0.032 $ & $  0.175 \pm 0.020 $ \\ 
20130918 &  8.9 & $ 1.190 \pm 0.023 $ & $  0.185 \pm 0.028 $ \\ 
20130918 & 13.9 & $ 1.295 \pm 0.014 $ & $  0.204 \pm 0.011 $ \\ 
20130918 & 21.1 & $ 1.336 \pm 0.013 $ & $  0.129 \pm 0.016 $ \\ 
20130918 & 32.0 & $ 1.227 \pm 0.031 $ & $ -0.172 \pm 0.031 $ \\ 
20130918 & 40.9 & $ 1.275 \pm 0.031 $ & $ -0.291 \pm 0.031 $ \\ 
20131026 &  1.5 & $ 0.588 \pm 0.037 $ & $ -0.055 \pm 0.150 $ \\ 
20131026 &  3.0 & $ 0.659 \pm 0.052 $ & $ -0.121 \pm 0.026 $ \\ 
20131026 &  5.4 & $ 0.826 \pm 0.020 $ & $ -0.015 \pm 0.023 $ \\ 
20131026 &  8.9 & $ 0.945 \pm 0.030 $ & $ -0.066 \pm 0.025 $ \\ 
20131026 & 13.9 & $ 1.085 \pm 0.015 $ & $ -0.016 \pm 0.008 $ \\ 
20131026 & 21.1 & $ 1.215 \pm 0.011 $ & $  0.022 \pm 0.003 $ \\ 
20131026 & 32.0 & $ 1.392 \pm 0.004 $ & $  0.057 \pm 0.007 $ \\ 
20131026 & 40.9 & $ 1.472 \pm 0.006 $ & $  0.013 \pm 0.014 $ \\ 
\tablebreak
20131129 &  1.5 & $ 0.683 \pm 0.025 $ & $ -0.055 \pm 0.197 $ \\ 
20131129 &  3.0 & $ 0.896 \pm 0.150 $ & $  0.137 \pm 0.049 $ \\ 
20131129 &  5.4 & $ 0.782 \pm 0.044 $ & $  0.147 \pm 0.026 $ \\ 
20131129 &  8.9 & $ 1.012 \pm 0.028 $ & $  0.241 \pm 0.018 $ \\ 
20131129 & 13.9 & $ 1.177 \pm 0.022 $ & $  0.335 \pm 0.008 $ \\ 
20131129 & 21.1 & $ 1.388 \pm 0.028 $ & $  0.393 \pm 0.017 $ \\ 
20131129 & 32.0 & $ 1.636 \pm 0.009 $ & $  0.492 \pm 0.003 $ \\ 
20131129 & 40.9 & $ 1.829 \pm 0.032 $ & $  0.499 \pm 0.005 $ \\ 
20131229 &  1.5 & \dots               & $  0.577 \pm 1.013 $ \\ 
20131229 &  3.0 & $ 0.729 \pm 0.043 $ & $ -0.138 \pm 0.052 $ \\ 
20131229 &  5.4 & $ 0.647 \pm 0.062 $ & $ -0.149 \pm 0.026 $ \\ 
20131229 &  8.9 & $ 0.791 \pm 0.025 $ & $ -0.242 \pm 0.018 $ \\ 
20131229 & 13.9 & $ 0.887 \pm 0.028 $ & $ -0.335 \pm 0.008 $ \\ 
20131229 & 21.1 & $ 0.999 \pm 0.013 $ & $ -0.392 \pm 0.016 $ \\ 
20131229 & 32.0 & $ 1.146 \pm 0.005 $ & $ -0.490 \pm 0.004 $ \\ 
20131229 & 40.9 & $ 1.334 \pm 0.054 $ & $ -0.498 \pm 0.005 $ \\ 
20140215 &  1.5 & $ 0.538 \pm 0.095 $ & $ -0.032 \pm 0.189 $ \\ 
20140215 &  3.0 & $ 0.680 \pm 0.033 $ & $ -0.054 \pm 0.032 $ \\ 
20140215 &  5.4 & $ 0.720 \pm 0.040 $ & $ -0.043 \pm 0.023 $ \\ 
20140215 &  8.9 & $ 0.877 \pm 0.022 $ & $ -0.010 \pm 0.020 $ \\ 
20140215 & 13.9 & $ 0.970 \pm 0.012 $ & $ -0.038 \pm 0.009 $ \\ 
20140215 & 21.1 & $ 1.135 \pm 0.008 $ & $ -0.027 \pm 0.009 $ \\ 
20140215 & 32.0 & $ 1.374 \pm 0.003 $ & $  0.035 \pm 0.007 $ \\ 
20140215 & 40.9 & $ 1.556 \pm 0.015 $ & $  0.080 \pm 0.014 $ \\ 
\tablebreak
20140322 &  1.5 & $ 0.597 \pm 0.056 $ & $  0.009 \pm 0.049 $ \\ 
20140322 &  3.0 & $ 0.682 \pm 0.025 $ & $ -0.056 \pm 0.011 $ \\ 
20140322 &  5.4 & $ 0.999 \pm 0.042 $ & $ -0.065 \pm 0.019 $ \\ 
20140322 &  8.9 & $ 0.860 \pm 0.017 $ & $ -0.042 \pm 0.013 $ \\ 
20140322 & 13.9 & $ 1.001 \pm 0.015 $ & $ -0.020 \pm 0.005 $ \\ 
20140322 & 21.1 & $ 1.166 \pm 0.008 $ & $ -0.003 \pm 0.013 $ \\ 
20140322 & 32.0 & $ 1.392 \pm 0.002 $ & $  0.042 \pm 0.004 $ \\ 
20140322 & 40.9 & $ 1.526 \pm 0.004 $ & $  0.050 \pm 0.013 $ \\ 
20140426 &  1.5 & $ 0.512 \pm 0.077 $ & $ -0.076 \pm 0.219 $ \\ 
20140426 &  3.0 & $ 0.775 \pm 0.039 $ & $  0.039 \pm 0.017 $ \\ 
20140426 &  5.4 & $ 0.776 \pm 0.068 $ & $  0.007 \pm 0.034 $ \\ 
20140426 &  8.9 & $ 0.862 \pm 0.011 $ & $ -0.016 \pm 0.029 $ \\ 
20140426 & 13.9 & $ 0.944 \pm 0.009 $ & $ -0.088 \pm 0.005 $ \\ 
20140426 & 21.1 & $ 1.057 \pm 0.005 $ & $ -0.122 \pm 0.004 $ \\ 
20140426 & 32.0 & $ 1.264 \pm 0.006 $ & $ -0.124 \pm 0.006 $ \\ 
20140426 & 40.9 & $ 1.466 \pm 0.005 $ & $ -0.071 \pm 0.007 $ \\ 
20140531 &  1.5 & $ 0.590 \pm 0.009 $ & $  0.045 \pm 0.034 $ \\ 
20140531 &  3.0 & $ 0.734 \pm 0.035 $ & $ -0.000 \pm 0.015 $ \\ 
20140531 &  5.4 & $ 0.828 \pm 0.043 $ & $ -0.052 \pm 0.020 $ \\ 
20140531 &  8.9 & $ 0.891 \pm 0.015 $ & $ -0.089 \pm 0.005 $ \\ 
20140531 & 13.9 & $ 1.019 \pm 0.011 $ & $ -0.065 \pm 0.008 $ \\ 
20140531 & 21.1 & $ 1.166 \pm 0.002 $ & $ -0.030 \pm 0.001 $ \\ 
20140531 & 32.0 & $ 1.428 \pm 0.058 $ & $  0.044 \pm 0.008 $ \\ 
20140531 & 40.9 & $ 1.871 \pm 0.177 $ & $  0.112 \pm 0.008 $ \\ 
\enddata
\end{deluxetable}

\begin{deluxetable}{llrrl}
\tablecaption{ALMA Observations \label{tab:almaobs}}
\tablehead{
\colhead{Epoch} & \colhead{UT} & \colhead{Beam B6 (230 GHz)} & \colhead{Beam B7 (345 GHz)} & \colhead{Flux Cal.} \\
                &              & (arcsec$^2$, deg) & (arcsec$^2$, deg) \\ 
}
\startdata
20130708 &  02:14-02:44 & $1.14 \times 0.53$,  65.1 & $0.38 \times 0.29$,  73.8 & Titan \\
20131102 &  22:39-22:57 & $0.91 \times 0.46$, -84.2 & $0.51 \times 0.30$, -87.2 & Neptune \\
20131218 &  14:07-14:27 & $1.36 \times 0.64$,  88.3 & $0.93 \times 0.44$,  86.0 & Titan \\
20140128 &  10:26-10:44 & $2.39 \times 0.78$,  78.8 & $1.64 \times 0.52$,  77.6 & Titan \\
20140222 &  09:20-09:45 & $1.89 \times 0.92$,  69.3 & $1.28 \times 0.62$,  68.2 & Titan \\
20140321 &  09:17-09:35 & $0.99 \times 0.67$,  79.3 & $0.65 \times 0.43$,  77.1 & Titan \\
20140425 &  06:59-07:29 & $0.71 \times 0.58$,  78.7 & $0.45 \times 0.38$,  88.8 & Titan \\
20140527 &  05:47-06:06 & $0.62 \times 0.50$, -75.2 & $0.40 \times 0.33$, -75.8 & Titan \\
\enddata
\end{deluxetable}

\begin{deluxetable}{lrrrrrrrrr}
\rotate
\tabletypesize{\scriptsize}
\tablecaption{ALMA Results \label{tab:almaresults}}
\tablehead{
\colhead{Epoch} &
\colhead{Source} &
\multicolumn{8}{c}{Flux Density (Jy)} \\
&
&
\colhead{218.0 GHz} &
\colhead{220.0 GHz} &
\colhead{231.9 GHz} &
\colhead{233.8 GHz} &
\colhead{341.6 GHz} &
\colhead{343.6 GHz} &
\colhead{351.7 GHz} &
\colhead{353.6 GHz} \\
}
\startdata
20130708 & SgrAStar & 3.792 & 3.795 & 3.872 & 3.892 & 4.259 & 4.278 & 4.270 & 4.265 \\ 
20131102 & SgrAStar & 4.635 & 4.649 & 4.752 & 4.758 & 4.537 & 4.564 & 4.569 & 4.537 \\ 
20131218 & SgrAStar & 2.976 & 2.970 & 2.997 & 3.010 & 3.032 & 3.046 & 3.012 & 3.003 \\ 
20140128 & SgrAStar & 3.868 & 3.868 & 3.944 & 3.957 & 4.293 & 4.313 & 4.309 & 4.294 \\ 
20140222 & SgrAStar & 2.811 & 2.803 & 2.800 & 2.801 & 2.332 & 2.341 & 2.311 & 2.301 \\ 
20140321 & SgrAStar & 4.078 & 4.047 & 3.983 & 4.002 & 3.476 & 3.447 & 3.428 & 3.406 \\ 
20140425 & SgrAStar & 4.121 & 4.111 & 4.004 & 4.163 & 4.296 & 4.287 & 4.295 & 4.064 \\ 
20140527 & SgrAStar & 3.054 & 3.048 & 3.054 & 3.051 & 2.587 & 2.596 & 2.568 & 2.556 \\ 
20130708 & J1733-130 & 1.359 & 1.337 & 1.256 & 1.264 & 0.952 & 0.951 & 0.917 & 0.857 \\ 
20131102 & J1733-130 & 1.586 & 1.570 & 1.429 & 1.413 & 0.998 & 0.994 & 0.995 & 0.998 \\ 
20131218 & J1733-130 & 1.771 & 1.754 & 1.668 & 1.657 & 1.311 & 1.273 & 1.261 & 1.263 \\ 
20140128 & J1733-130 & 1.576 & 1.561 & 1.477 & 1.482 & 1.203 & 1.190 & 1.181 & 1.173 \\ 
20140222 & J1733-130 & 1.649 & 1.638 & 1.557 & 1.560 & 1.214 & 1.201 & 1.192 & 1.187 \\ 
20140321 & J1733-130 & 1.517 & 1.502 & 1.426 & 1.426 & 1.125 & 1.109 & 1.097 & 1.092 \\ 
20140425 & J1733-130 & 1.545 & 1.532 & 1.406 & 1.449 & 1.140 & 1.126 & 1.114 & 1.053 \\ 
20140527 & J1733-130 & 1.619 & 1.600 & 1.517 & 1.514 & 1.179 & 1.169 & 1.142 & 1.150 \\ 
20131218 & J1427-421 & 2.899 & 2.876 & 2.739 & 2.750 & 2.131 & 2.079 & 2.065 & 2.058 \\ 
20140128 & J1427-421 & 3.073 & 3.041 & 2.908 & 2.907 & 2.422 & 2.387 & 2.375 & 2.379 \\ 
20140222 & J1427-421 & 3.370 & 3.353 & 3.201 & 3.206 & 2.665 & 2.658 & 2.639 & 2.645 \\ 
20140321 & J1427-421 & 3.773 & 3.735 & 3.570 & 3.582 & 2.931 & 2.900 & 2.859 & 2.856 \\ 
20140425 & J1427-421 & 4.015 & 3.984 & 3.669 & 3.799 & -0.019 & -0.019 & -0.016 & -0.015 \\ 
20140527 & J1427-421 & 4.227 & 4.187 & 4.002 & 4.002 & 3.253 & 3.217 & 3.191 & 3.176 \\ 
20130708 & J1700-261 & 0.967 & 0.962 & 0.914 & 0.916 & 0.701 & 0.685 & 0.666 & 0.651 \\ 
20131102 & J1700-261 & 0.791 & 0.784 & 0.721 & 0.720 & 0.517 & 0.518 & 0.525 & 0.522 \\ 
20131218 & J1700-261 & 0.774 & 0.775 & 0.732 & 0.734 & 0.592 & 0.583 & 0.574 & 0.585 \\ 
20140128 & J1700-261 & 0.887 & 0.882 & 0.842 & 0.846 & 0.716 & 0.707 & 0.706 & 0.700 \\ 
20140222 & J1700-261 & 0.969 & 0.961 & 0.921 & 0.924 & 0.749 & 0.744 & 0.738 & 0.738 \\ 
20140321 & J1700-261 & 0.846 & 0.838 & 0.800 & 0.804 & 0.664 & 0.657 & 0.650 & 0.646 \\ 
20140425 & J1700-261 & \dots & \dots & \dots & \dots & 0.773 & 0.763 & 0.757 & 0.723 \\ 
20140527 & J1700-261 & 1.072 & 1.062 & 1.004 & 1.005 & 0.824 & 0.799 & 0.803 & 0.799 \\ 
\enddata
\end{deluxetable}

\begin{deluxetable}{lrrrr}
\tablecaption{SMA Results \label{tab:smaresults}}
\tablehead{
\colhead{Epoch} &
\colhead{UT} &
\colhead{Frequency} &
\colhead{$S_{\rm \sgra}$} &
\colhead{$S_{\rm J1733-130}$} 
\\
&
&
\colhead{(GHz)} &
\colhead{(Jy)} &
\colhead{(Jy)} 
}
\startdata
20130601 & 13:18 & 218.60 & $ 3.217 \pm 0.168 $ & $ 1.514 \pm 0.083 $ \\ 
20130605 & 13:23 & 336.40 & $ 2.680 \pm 0.145 $ & $ 1.110 \pm 0.081 $ \\ 
20130609 & 13:21 & 271.80 & $ 4.743 \pm 0.246 $ & $ 1.276 \pm 0.081 $ \\ 
20130610 & 13:32 & 265.99 & $ 3.237 \pm 0.188 $ & $ 1.271 \pm 0.106 $ \\ 
20130612 & 12:15 & 271.79 & $ 4.055 \pm 0.206 $ & $ 1.319 \pm 0.078 $ \\ 
20130615 & 12:08 & 218.48 & $ 3.951 \pm 0.210 $ & $ 1.514 \pm 0.086 $ \\ 
20130616 & 12:53 & 218.95 & $ 3.956 \pm 0.210 $ & $ 1.452 \pm 0.088 $ \\ 
20130617 & 12:15 & 219.71 & $ 4.836 \pm 0.245 $ & $ 1.452 \pm 0.080 $ \\ 
20130618 & 12:03 & 218.47 & $ 4.605 \pm 0.237 $ & $ 1.605 \pm 0.098 $ \\ 
20130619 & 12:41 & 218.95 & $ 3.618 \pm 0.193 $ & $ 1.489 \pm 0.088 $ \\ 
20130620 & 12:13 & 214.60 & $ 3.982 \pm 0.207 $ & $ 1.451 \pm 0.097 $ \\ 
20130621 & 12:37 & 218.77 & $ 4.018 \pm 0.213 $ & $ 1.454 \pm 0.119 $ \\ 
20130622 & 12:41 & 218.93 & $ 3.957 \pm 0.205 $ & $ 1.451 \pm 0.086 $ \\ 
20130623 & 11:39 & 218.68 & $ 4.079 \pm 0.249 $ & $ 1.464 \pm 0.114 $ \\ 
20130624 & 12:10 & 218.75 & $ 3.986 \pm 0.224 $ & $ 1.356 \pm 0.141 $ \\ 
20130626 & 11:43 & 218.68 & $ 3.980 \pm 0.211 $ & $ 1.395 \pm 0.083 $ \\ 
20130630 & 12:13 & 333.40 & $ 3.832 \pm 0.259 $ & $ 1.053 \pm 0.137 $ \\ 
20130701 & 10:35 & 342.96 & $ 4.023 \pm 0.212 $ & $ 1.056 \pm 0.083 $ \\ 
20130705 & 07:18 & 226.86 & $ 4.204 \pm 0.228 $ & $ 1.384 \pm 0.073 $ \\ 
20130705 & 08:55 & 226.86 & $ 4.796 \pm 0.245 $ & $ 1.420 \pm 0.072 $ \\ 
20130705 & 11:41 & 226.86 & $ 3.980 \pm 0.204 $ & $ 1.398 \pm 0.073 $ \\ 
20130708 & 11:07 & 218.87 & $ 3.989 \pm 0.205 $ & $ 1.409 \pm 0.074 $ \\ 
20130713 & 10:55 & 336.38 & $ 3.524 \pm 0.195 $ & $ 1.074 \pm 0.067 $ \\ 
20130714 & 10:44 & 333.40 & $ 3.267 \pm 0.194 $ & $ 1.116 \pm 0.101 $ \\ 
20130716 & 10:36 & 218.84 & $ 3.265 \pm 0.184 $ & $ 1.430 \pm 0.105 $ \\ 
20130717 & 11:05 & 265.99 & $ 3.439 \pm 0.181 $ & $ 1.263 \pm 0.088 $ \\ 
20130719 & 10:32 & 265.97 & $ 3.315 \pm 0.170 $ & $ 1.090 \pm 0.064 $ \\ 
20130720 & 10:42 & 271.78 & $ 3.597 \pm 0.188 $ & $ 1.206 \pm 0.080 $ \\ 
20130721 & 10:30 & 218.69 & $ 2.887 \pm 0.152 $ & $ 1.267 \pm 0.064 $ \\ 
20130723 & 10:43 & 218.84 & $ 3.463 \pm 0.215 $ & $ 1.391 \pm 0.141 $ \\ 
20130724 & 10:45 & 218.90 & $ 3.522 \pm 0.207 $ & $ 1.406 \pm 0.098 $ \\ 
20130728 & 10:12 & 336.07 & $ 3.467 \pm 0.268 $ & $ 1.174 \pm 0.258 $ \\ 
20130731 & 10:07 & 218.79 & $ 3.869 \pm 0.401 $ & $ 1.347 \pm 0.140 $ \\ 
20130801 & 09:58 & 218.79 & $ 3.475 \pm 0.248 $ & $ 1.419 \pm 0.085 $ \\ 
20130802 & 09:38 & 218.79 & $ 4.200 \pm 0.217 $ & $ 1.454 \pm 0.080 $ \\ 
20130803 & 09:01 & 218.82 & $ 4.740 \pm 0.244 $ & $ 1.428 \pm 0.072 $ \\ 
20130804 & 08:46 & 334.73 & $ 5.044 \pm 0.287 $ & $ 1.122 \pm 0.177 $ \\ 
20130804 & 09:39 & 215.44 & $ 4.727 \pm 0.250 $ & $ 1.499 \pm 0.089 $ \\ 
20130805 & 08:51 & 336.98 & $ 4.854 \pm 0.340 $ & $ 1.151 \pm 0.350 $ \\ 
20130806 & 09:49 & 218.96 & $ 3.788 \pm 0.205 $ & $ 1.558 \pm 0.087 $ \\ 
20130808 & 09:09 & 333.98 & $ 4.014 \pm 0.291 $ & $ 1.091 \pm 0.122 $ \\ 
20130809 & 08:34 & 218.85 & $ 4.298 \pm 0.224 $ & $ 1.538 \pm 0.098 $ \\ 
20130810 & 09:13 & 218.77 & $ 4.356 \pm 0.333 $ & $ 1.528 \pm 0.162 $ \\ 
20130811 & 08:55 & 218.77 & $ 3.857 \pm 0.294 $ & $ 1.451 \pm 0.138 $ \\ 
20130812 & 09:10 & 218.99 & $ 3.744 \pm 0.205 $ & $ 1.557 \pm 0.096 $ \\ 
20130813 & 09:10 & 218.66 & $ 4.642 \pm 0.246 $ & $ 1.513 \pm 0.089 $ \\ 
20130814 & 08:19 & 334.00 & $ 4.469 \pm 0.275 $ & $ 1.109 \pm 0.124 $ \\ 
20130815 & 03:27 & 226.85 & $ 3.307 \pm 0.172 $ & $ 1.517 \pm 0.085 $ \\ 
20130815 & 03:56 & 226.85 & $ 3.228 \pm 0.173 $ & $ 1.475 \pm 0.081 $ \\ 
20130815 & 04:27 & 226.85 & $ 3.114 \pm 0.163 $ & $ 1.514 \pm 0.084 $ \\ 
20130815 & 04:56 & 226.85 & $ 3.062 \pm 0.165 $ & $ 1.523 \pm 0.084 $ \\ 
20130815 & 05:27 & 226.85 & $ 3.118 \pm 0.164 $ & $ 1.510 \pm 0.083 $ \\ 
20130815 & 05:56 & 226.85 & $ 3.641 \pm 0.192 $ & $ 1.490 \pm 0.078 $ \\ 
20130815 & 06:25 & 226.85 & $ 3.623 \pm 0.186 $ & $ 1.510 \pm 0.081 $ \\ 
20130815 & 06:57 & 226.85 & $ 3.582 \pm 0.184 $ & $ 1.497 \pm 0.080 $ \\ 
20130815 & 07:25 & 226.85 & $ 3.638 \pm 0.188 $ & $ 1.516 \pm 0.088 $ \\ 
20130815 & 08:01 & 226.85 & $ 3.742 \pm 0.196 $ & $ 1.494 \pm 0.078 $ \\ 
20130815 & 08:26 & 226.85 & $ 3.533 \pm 0.182 $ & $ 1.495 \pm 0.083 $ \\ 
20130815 & 08:53 & 226.85 & $ 3.529 \pm 0.182 $ & $ 1.497 \pm 0.081 $ \\ 
20130815 & 09:24 & 226.85 & $ 3.332 \pm 0.176 $ & $ 1.530 \pm 0.090 $ \\ 
20130816 & 08:12 & 220.59 & $ 3.149 \pm 0.165 $ & $ 1.475 \pm 0.082 $ \\ 
20130914 & 06:24 & 218.93 & $ 4.685 \pm 0.246 $ & $ 1.540 \pm 0.083 $ \\ 
20130916 & 06:28 & 331.32 & $ 5.048 \pm 0.282 $ & $ 1.100 \pm 0.114 $ \\ 
20130917 & 06:43 & 219.19 & $ 4.736 \pm 0.238 $ & $ 1.550 \pm 0.079 $ \\ 
20130918 & 06:15 & 331.31 & $ 5.492 \pm 0.280 $ & $ 1.149 \pm 0.091 $ \\ 
20130920 & 06:12 & 331.31 & $ 5.310 \pm 0.318 $ & $ 1.162 \pm 0.169 $ \\ 
20130921 & 05:01 & 219.87 & $ 4.778 \pm 0.242 $ & $ 1.496 \pm 0.083 $ \\ 
20130921 & 05:01 & 334.79 & $ 5.751 \pm 0.297 $ & $ 1.155 \pm 0.104 $ \\ 
20130922 & 05:49 & 220.64 & $ 4.566 \pm 0.233 $ & $ 1.513 \pm 0.087 $ \\ 
20130922 & 05:49 & 355.92 & $ 6.090 \pm 0.342 $ & $ 1.132 \pm 0.197 $ \\ 
20130924 & 05:38 & 331.37 & $ 5.093 \pm 0.297 $ & $ 1.111 \pm 0.161 $ \\ 
20130925 & 06:30 & 219.21 & $ 4.271 \pm 0.216 $ & $ 1.540 \pm 0.080 $ \\ 
20130926 & 06:28 & 218.93 & $ 4.345 \pm 0.219 $ & $ 1.548 \pm 0.081 $ \\ 
20130928 & 05:24 & 220.71 & $ 4.229 \pm 0.214 $ & $ 1.561 \pm 0.086 $ \\ 
20130928 & 05:24 & 355.92 & $ 5.330 \pm 0.294 $ & $ 1.092 \pm 0.153 $ \\ 
20130930 & 05:54 & 220.28 & $ 4.335 \pm 0.219 $ & $ 1.557 \pm 0.082 $ \\ 
20131002 & 05:32 & 220.27 & $ 5.007 \pm 0.261 $ & $ 1.517 \pm 0.095 $ \\ 
20131003 & 05:30 & 220.27 & $ 3.438 \pm 0.190 $ & $ 1.577 \pm 0.125 $ \\ 
20131004 & 04:22 & 219.85 & $ 5.101 \pm 0.260 $ & $ 1.607 \pm 0.090 $ \\ 
20131006 & 05:06 & 334.92 & $ 4.424 \pm 0.244 $ & $ 1.163 \pm 0.113 $ \\ 
20131007 & 05:00 & 220.27 & $ 4.574 \pm 0.232 $ & $ 1.557 \pm 0.087 $ \\ 
20131008 & 04:37 & 239.61 & $ 4.035 \pm 0.215 $ & $ 1.521 \pm 0.085 $ \\ 
20131012 & 04:26 & 218.83 & $ 5.420 \pm 0.290 $ & $ 1.555 \pm 0.095 $ \\ 
20131013 & 04:27 & 218.83 & $ 4.530 \pm 0.235 $ & $ 1.492 \pm 0.090 $ \\ 
20131022 & 04:08 & 239.61 & $ 3.630 \pm 0.190 $ & $ 1.536 \pm 0.098 $ \\ 
20131101 & 03:25 & 218.90 & $ 4.532 \pm 0.235 $ & $ 1.618 \pm 0.098 $ \\ 
20131102 & 03:16 & 219.37 & $ 4.118 \pm 0.219 $ & $ 1.585 \pm 0.087 $ \\ 
20140116 & 19:58 & 225.02 & $ 2.957 \pm 0.150 $ & $ 1.383 \pm 0.071 $ \\ 
20140408 & 13:59 & 266.10 & $ 3.482 \pm 0.180 $ & $ 1.271 \pm 0.070 $ \\ 
20140409 & 16:42 & 334.01 & $ 3.232 \pm 0.183 $ & $ 1.157 \pm 0.106 $ \\ 
20140410 & 17:01 & 219.74 & $ 4.211 \pm 0.214 $ & $ 1.473 \pm 0.082 $ \\ 
20140412 & 16:41 & 218.92 & $ 4.361 \pm 0.221 $ & $ 1.549 \pm 0.081 $ \\ 
20140415 & 16:22 & 356.05 & $ 4.146 \pm 0.280 $ & $ 1.162 \pm 0.213 $ \\ 
20140416 & 16:40 & 218.78 & $ 3.820 \pm 0.192 $ & $ 1.441 \pm 0.075 $ \\ 
20140417 & 16:37 & 218.43 & $ 3.976 \pm 0.203 $ & $ 1.567 \pm 0.082 $ \\ 
20140422 & 16:10 & 220.56 & $ 3.228 \pm 0.202 $ & $ 1.491 \pm 0.091 $ \\ 
20140423 & 16:47 & 218.97 & $ 3.479 \pm 0.188 $ & $ 1.497 \pm 0.095 $ \\ 
20140425 & 15:19 & 215.27 & $ 3.301 \pm 0.179 $ & $ 1.578 \pm 0.083 $ \\ 
20140505 & 15:01 & 219.02 & $ 4.022 \pm 0.212 $ & $ 1.575 \pm 0.083 $ \\ 
20140506 & 14:49 & 218.74 & $ 3.409 \pm 0.176 $ & $ 1.588 \pm 0.087 $ \\ 
20140510 & 13:43 & 336.95 & $ 4.499 \pm 0.237 $ & $ 1.253 \pm 0.100 $ \\ 
20140511 & 14:15 & 336.95 & $ 3.387 \pm 0.182 $ & $ 1.241 \pm 0.112 $ \\ 
20140512 & 15:43 & 218.93 & $ 4.155 \pm 0.214 $ & $ 1.648 \pm 0.089 $ \\ 
20140521 & 14:31 & 219.00 & $ 3.368 \pm 0.181 $ & $ 1.584 \pm 0.087 $ \\ 
20140522 & 14:12 & 219.00 & $ 4.127 \pm 0.215 $ & $ 1.571 \pm 0.084 $ \\ 
20140523 & 14:15 & 332.14 & $ 1.999 \pm 0.135 $ & $ 1.205 \pm 0.113 $ \\ 
20140826 & 08:13 & 335.55 & $ 3.173 \pm 0.168 $ & $ 0.976 \pm 0.070 $ \\ 
20140829 & 07:09 & 353.75 & $ 4.458 \pm 0.271 $ & $ 0.880 \pm 0.208 $ \\ 
20140902 & 07:39 & 241.36 & $ 2.955 \pm 0.153 $ & $ 1.156 \pm 0.069 $ \\ 
20140903 & 07:45 & 271.52 & $ 2.799 \pm 0.145 $ & $ 1.054 \pm 0.069 $ \\ 
20140904 & 07:07 & 220.77 & $ 2.586 \pm 0.138 $ & $ 1.259 \pm 0.082 $ \\ 
20140905 & 07:20 & 220.77 & $ 2.924 \pm 0.154 $ & $ 1.242 \pm 0.082 $ \\ 
20140909 & 07:18 & 241.36 & $ 3.020 \pm 0.176 $ & $ 1.149 \pm 0.092 $ \\ 
20140912 & 06:06 & 335.59 & $ 2.421 \pm 0.136 $ & $ 0.750 \pm 0.068 $ \\ 
20140915 & 05:45 & 220.20 & $ 2.158 \pm 0.153 $ & $ 1.165 \pm 0.107 $ \\ 
20140917 & 06:19 & 333.39 & $ 2.500 \pm 0.141 $ & $ 0.670 \pm 0.084 $ \\ 
20140925 & 05:40 & 214.17 & $ 2.154 \pm 0.119 $ & $ 1.181 \pm 0.066 $ \\ 
20140930 & 05:15 & 219.42 & $ 2.309 \pm 0.142 $ & $ 1.159 \pm 0.066 $ \\ 
20141001 & 05:50 & 213.23 & $ 2.747 \pm 0.159 $ & $ 1.195 \pm 0.069 $ \\ 
20141002 & 05:05 & 213.44 & $ 3.456 \pm 0.182 $ & $ 1.225 \pm 0.068 $ \\ 
20141003 & 05:39 & 219.52 & $ 3.191 \pm 0.178 $ & $ 1.196 \pm 0.069 $ \\ 
20141005 & 05:25 & 218.68 & $ 2.557 \pm 0.160 $ & $ 1.262 \pm 0.075 $ \\ 
20141006 & 05:36 & 218.67 & $ 2.659 \pm 0.167 $ & $ 1.190 \pm 0.075 $ \\ 
20141009 & 05:18 & 218.62 & $ 4.222 \pm 0.250 $ & $ 1.163 \pm 0.070 $ \\ 
20141015 & 05:24 & 219.71 & $ 3.514 \pm 0.317 $ & $ 1.149 \pm 0.140 $ \\ 
20141016 & 05:16 & 218.90 & $ 3.804 \pm 0.259 $ & $ 1.148 \pm 0.087 $ \\ 
20141021 & 05:03 & 220.26 & $ 4.190 \pm 0.235 $ & $ 1.175 \pm 0.090 $ \\ 
20141025 & 04:09 & 219.42 & $ 3.139 \pm 0.170 $ & $ 1.150 \pm 0.065 $ \\ 
20141029 & 04:10 & 335.58 & $ 3.783 \pm 0.227 $ & $ 0.824 \pm 0.115 $ \\ 
20141030 & 04:03 & 335.58 & $ 3.365 \pm 0.219 $ & $ 0.893 \pm 0.130 $ \\ 
20141031 & 04:18 & 356.16 & $ 2.961 \pm 0.282 $ & $ 0.971 \pm 0.178 $ \\ 
20141103 & 04:10 & 220.27 & $ 3.502 \pm 0.316 $ & $ 1.158 \pm 0.175 $ \\ 
20141104 & 04:09 & 218.63 & $ 3.701 \pm 0.346 $ & $ 1.062 \pm 0.251 $ \\ 
20141107 & 03:49 & 220.26 & $ 3.680 \pm 0.281 $ & $ 1.102 \pm 0.096 $ \\ 
20141109 & 03:55 & 333.33 & $ 3.531 \pm 0.252 $ & $ 0.672 \pm 0.117 $ \\ 
20141110 & 03:46 & 333.36 & $ 2.359 \pm 0.199 $ & $ 0.787 \pm 0.084 $ \\ 
20141112 & 03:51 & 333.36 & $ 2.609 \pm 0.201 $ & $ 0.717 \pm 0.093 $ \\ 
\enddata
\end{deluxetable}

\begin{deluxetable}{rrrrrrrr}
\tabletypesize{\scriptsize}
\tablecaption{Mean Spectrum of \sgra\ \label{tab:meanspec}}
\tablehead{
\colhead{Freq.} &
\colhead{Mean Flux Density} &
\colhead{Standard Dev.} &
\colhead{Minimum Flux  Density} &
\colhead{Maximum Flux Density} &
\colhead{$N_{\rm epoch}$} &
\colhead{Tel.} \\
\colhead{(GHz)} & \colhead{(Jy)} & \colhead{(Jy)} & \colhead{(Jy)} & \colhead{(Jy)} 
}
\startdata
  1.6 & 0.592 & 0.028 & 0.512 & 0.683 &  8 & VLA \\ 
  3.1 & 0.702 & 0.032 & 0.659 & 0.896 & 10 & VLA \\ 
  5.4 & 0.870 & 0.118 & 0.647 & 1.134 & 12 & VLA \\ 
  9.0 & 0.932 & 0.129 & 0.791 & 1.276 & 12 & VLA \\ 
 14.0 & 1.075 & 0.135 & 0.887 & 1.749 & 14 & VLA \\ 
 21.1 & 1.164 & 0.052 & 0.999 & 1.562 & 15 & VLA \\ 
 32.0 & 1.382 & 0.087 & 1.146 & 1.648 & 15 & VLA \\ 
 40.9 & 1.485 & 0.073 & 1.275 & 1.882 & 15 & VLA \\ 
218.0 & 3.667 & 0.650 & 2.811 & 4.635 &  8 & ALMA \\ 
220.0 & 3.661 & 0.652 & 2.803 & 4.649 &  8 & ALMA \\ 
231.9 & 3.676 & 0.664 & 2.800 & 4.752 &  8 & ALMA \\ 
233.8 & 3.704 & 0.680 & 2.801 & 4.758 &  8 & ALMA \\ 
341.6 & 3.602 & 0.866 & 2.332 & 4.537 &  8 & ALMA \\ 
343.6 & 3.609 & 0.870 & 2.341 & 4.564 &  8 & ALMA \\ 
351.7 & 3.595 & 0.884 & 2.311 & 4.569 &  8 & ALMA \\ 
353.6 & 3.553 & 0.860 & 2.301 & 4.537 &  8 & ALMA \\ 
216.8 & 3.677 & 0.762 & 2.154 & 5.420 & 68 & SMA \\ 
223.9 & 3.391 & 0.489 & 2.586 & 4.796 & 23 & SMA \\ 
238.2 & 3.310 & 0.424 & 2.955 & 4.035 &  4 & SMA \\ 
266.8 & 3.369 & 0.096 & 3.237 & 3.482 &  4 & SMA \\ 
274.0 & 3.526 & 0.697 & 2.799 & 4.743 &  4 & SMA \\ 
331.1 & 3.205 & 1.074 & 1.999 & 5.492 & 14 & SMA \\ 
338.3 & 3.436 & 0.863 & 2.421 & 5.751 & 13 & SMA \\ 
352.6 & 4.890 & 0.721 & 4.146 & 6.090 &  4 & SMA \\ 
\enddata
\end{deluxetable}

\end{document}